\documentclass[%
 reprint,showpacs,
 amsmath,amssymb,
 aps,nofootinbib,superscriptaddress
]{revtex4-1}
 \usepackage{ulem}
    \usepackage{float}
\usepackage{graphicx}
\usepackage{dcolumn}
\usepackage[colorlinks=true, linkcolor = red, citecolor = blue]{hyperref}
\usepackage{subcaption}
\usepackage{float}
\usepackage{xcolor}

\begin{document}
\newcommand{\lu}[1]{\textcolor{red}{#1}}
\newcommand{\trd}[1]{\textcolor{black}{(trd: #1)}\color{black}}

\title{Core-Halo Mass Relation in Scalar Field Dark Matter Models and its Consequences for the Formation of Supermassive Black Holes}

\author{Luis E. Padilla}
\email{epadilla@fis.cinvestav.mx}
\affiliation{Departamento de F\'isica, Centro de Investigaci\'on y de Estudios Avanzados del IPN, A.P. 14-740, 07000 M\'exico D.F.,
  M\'exico.}
\affiliation{Instituto de Ciencias F\'isicas, Universidad Nacional Aut\'onoma de M\'exico,
Apdo. Postal 48-3, 62251 Cuernavaca, Morelos, M\'exico.}
\affiliation{Mesoamerican Centre for Theoretical Physics,
Universidad Aut\'onoma de Chiapas, Carretera Zapata Km. 4, Real
del Bosque (Ter\'an), Tuxtla Guti\'errez 29040, Chiapas, M\'exico}

  \author{Tanja Rindler-Daller}
  \email{tanja.rindler-daller@univie.ac.at}
\affiliation{Institut f\"ur Astrophysik, Universit\"atssternwarte Wien, University of Vienna, T\"urkenschanzstr. 17, 1180 Vienna, Austria}
  \author{Paul R. Shapiro}
  \email{shapiro@astro.as.utexas.edu}
  \affiliation{Department of Astronomy and Texas Cosmology Center, The University of Texas at Austin, 2515 Speedway C1400, Austin, TX 78712, USA}
\author{Tonatiuh Matos}
\email{tonatiuh.matos@cinvestav.mx}
\affiliation{Departamento de F\'isica, Centro de Investigaci\'on y de Estudios Avanzados del IPN, A.P. 14-740, 07000 M\'exico D.F.,
  M\'exico.}
    \author{J. Alberto V\'azquez}
  \email{javazquez@icf.unam.mx}
  \affiliation{Instituto de Ciencias F\'isicas, Universidad Nacional Aut\'onoma de M\'exico, 
Apdo. Postal 48-3, 62251 Cuernavaca, Morelos, M\'exico.}

\date{\today}

\begin{abstract}

Scalar-field dark matter (SFDM) halos exhibit a core-envelope structure with soliton-like cores and \textit{Cold Dark Matter} (CDM)-like envelopes. Simulations without self-interaction (free-field case) have reported a core-halo mass relation of the form $M_c\propto M_{h}^{\beta}$, with either  $\beta=1/3$ or $\beta=5/9$. These results can be understood if the core and halo follow some special energy or velocity scaling relations. We extend these core-halo mass relations here to include the case of SFDM with self-interaction, either repulsive or attractive, and investigate its implications for the possible gravitational instability and collapse of solitonic cores, leading to the formation of supermassive black holes (SMBHs). Core sizes are set by the larger of two length scales, the deBroglie wavelength (in free-field limit) or the radius $R_{TF}$ of the (n = 1)-polytrope for repulsive SFDM (in the Thomas-Fermi regime), depending upon particle mass $m$ and interaction strength $\lambda$. For parameters selected by previous literature to make $\sim$ Kpc-sized cores and CDM-like structure formation on large scales but suppressed  on small scales, we find that cores are stable for all galactic halos of interest, from the free-field to the repulsive TF limit. For attractive self-interaction in this regime, however, halos of mass $M_h\sim 10^{10}-10^{12} ~M_\odot$  have cores that collapse to form seed SMBHs with $M_{SMBH}\sim 10^{6}-10^8 M_\odot$, as  observations seem to require, while smaller-mass halos have stable cores, for particle masses $m = 2.14\times 10^{-22}-9.9\times 10^{-20}~ \rm{eV}/c^2$, if the free-field limit has $\beta=1/3$, or $m = 2.23\times 10^{-21}-1.7\times 10^{-18}~\rm{eV}/c^2$, if $\beta=5/9$. We also place bounds on $\lambda$ for this case.  For free-field and repulsive cases, if previous constraints on particle parameters are relaxed to allow much smaller (sub-galactic scale) cores, then halos can also form SMBHs, for the same range of halo and BH masses, as long as $\beta = 5/9$ is correct for the free-field limit.  In that case, structure formation in SFDM would be largely indistinguishable from that in CDM. As such, while these SFDM models might not help to resolve the small-scale structure problems of CDM, they would explain the formation of SMBHs quite naturally, which is otherwise not a direct feature of CDM. Since CDM, itself, has not yet been ruled out, such SFDM models must also be viable.
\end{abstract}

\maketitle


\section{\label{sec:intro}Introduction}

Two of the greatest puzzles in contemporary cosmology and fundamental physics are:
\begin{enumerate}
    \item The nature and origin of cosmic dark matter (DM).
    \item The origin of the SMBHs observed in galactic nuclei.
\end{enumerate}


Regarding the first point, the cosmological standard model $\Lambda$CDM suggests that DM is comprised of a nonrelativistic, collisionless gas -- CDM -- and usually assumed to be a weakly-interacting massive particle (WIMP) which originated as a thermal relic of the big bang \citep{cdm1,cdm2}. Although WIMP dark matter describes observations well at cosmological scales, it is in apparent conflict with some observations on small scales (e.g. the problem of cuspy-core halo density profiles, overproduction of satellite dwarfs within the Local Group, and others, see for example \citep{cdm3,cdm4,cdm5,cdm6,cdm7}). In addition, all attempts to detect WIMPs directly in the laboratory or indirectly by astronomical signals from their decay or annihilation in distant objects \citep{cdm8} have been unsuccessful, and a large range of particle parameters originally predicted to be detectable have thereby been ruled out. For this reason, it seems necessary to explore alternative models to standard $\Lambda$CDM that help us to solve all these issues. With this in mind, several models have been proposed, one of which considers that the DM is composed of an ultra-light real or complex scalar field, minimally coupled to gravity, and interacting only gravitationally with the rest of the matter as of a very early time in the cosmic evolution. 

The main idea of scalar fields as the DM in the Universe originated about two decades ago \citep{sf5,sf1,sf2,sf3,sf40,sf4,sf6,sf7}, although some hints can be traced further back in \citep{sf80,sf8}.  Since then the idea has been rediscovered by various authors with different names, for example: SFDM \citep{sf4}, fuzzy DM \citep{sf2}, wave DM \citep{sc4}, Bose-Einstein condensate DM \citep{sf10} or ultra-light axion DM \citep{sf11} (see also \citep{sf12}). However, its first systematic study started in \citep{sf13,sf14}. In this work, we choose to call the model ``SFDM".

{In order for the scalar field to behave as a ``cold" DM candidate, it is necessary that its Lagrangian possesses a quadratic term in the potential, 
\begin{equation}
    V(\varphi) = \frac{1}{2}\frac{m^2c^2}{\hbar^2}\varphi^2,
\end{equation}
which gives rise to a pressureless fluid behavior in the matter-dominated epoch of the Universe, where SFDM dominates all other cosmic components. } 

{The simplest models have only this term in the scalar field potential, i.e. there is one tunable parameter $m$, which is subject to constraints from observations, as we will describe below. Such models are usually termed ``fuzzy dark matter" 
-- we will often call it ``free SFDM/free field" or ``free case" in this paper.  However, we are interested to study more varied models by considering the addition of a further term in the Lagrangian of the form
\begin{equation}\label{SIpotential}
 V(\varphi)_{SI}=\frac{\lambda}{4\hbar c}\varphi^4.
\end{equation}
Similarly to the quadratic term, this quartic term may either stem directly from a fundamental particle description of SFDM, or it might result upon an expansion of a fundamental (even) potential, as e.g the cosine-type of ``axion-like" particles.  Self-interaction has been mostly neglected in previous literature, because of the smallness of the respective coupling parameter $\lambda$.  However, it is only recently that the community has embarked on studying models with self-interaction in more detail, because it turns out that self-interaction leads to qualitative differences, compared to the free case. }

{The astrophysical motivation to consider SFDM as a DM candidate has its root in the small-scale problems of CDM mentioned in the beginning. In order to reproduce galactic cores of order 1 kpc for the free SFDM case, the boson mass is typically assumed in the range of $m\sim 10^{-22}-10^{-20}~\rm{eV}/c^2$. However, once self-interaction is included, the boson mass can be much higher than those values, and yet produce large enough cores by tuning the ratio of $\lambda/m^4$. A re-assessment of the different constraints in the literature, as well as including the implications of our work here, will be presented in a later section (for a review of SFDM, see \citep{rev1,rev2,rev3,rev4,niemeyer2019small,RS}). Our work will focus on certain dynamical aspects of SFDM structure formation, namely the structure of equilibrium halos and possible implications for SMBH formation. 
}

Simulations of SFDM cores without self-interaction \citep{sc4,sc10,sc11,sc12,sc13,sc14,moczext1,moczext2} have shown that, upon multiple mergers, SFDM leads to cored density profiles in the inner region of galactic halos.
{These cores, referred to as ``solitons" in the literature \citep{sc14,sc15,sc16,sc17,sc4}, have been shown to have a size of order of the de Broglie wavelength of individual bosons,
\begin{equation}\label{xdb}
    \lambda_{dB}\propto \frac{1}{mv},
\end{equation}
where $v$ is the ``virial velocity" of the bosons, a result expected from analytic calculations. However, these cores have been found to be surrounded by an NFW-like envelope generated by quantum interference inherent to SFDM, following {a relation of the form:} 
\begin{equation}\label{mcmh}
M_c\propto M_h^{\beta}, 
\end{equation}
where $M_c$ and $M_h$ are the total core and halo mass, respectively. {The particular value of this $\beta$ parameter is still under debate given that different authors have obtained different results. On the one hand, in \citep{sc4,sc10} it was found from their fully cosmological simulations an expression that is well adjusted with a parameter $\beta = 1/3$. On the other hand, by adopting more simplified scenarios on galaxy formation but with better resolution, some authors \citep{moczext1,sc13} have found that a parameter $\beta = 5/9$ should describe correctly virialized core-halo mass structures in this scenario of SFDM.} This correlation between the halo core and its ``envelope" has not been anticipated by early work, though it is possible to understand the form of {each} correlation in an \textit{a posteriori} way, using analytic arguments. Indeed, the fact that these correlations have been established by simulations,} 
offers a unique opportunity to understand and extend the correlations by considering novel physical effects such as the addition of self-interaction.

{The core-envelope structure of SFDM halos with self-interaction has not yet been established by 3D cosmological simulations, and yet we expect such a structure, as well. Preliminary results of 1D simulations show that a core-envelope structure also arises in the strongly self-interacting regime of SFDM, the so-called Thomas-Fermi (TF) regime (T.A.Dawoodbhoy, P.R.Shapiro, T.Rindler-Daller, to be subm.). This is good news for SFDM, because the (quantum) cores alone cannot explain the big range of galactic halo masses found in the Universe, and it is this simple observation which ``mandates" that cores (with or without self-interaction) have to be enshrouded by some envelopes, if SFDM is regarded as an alternative to CDM. Early indications of the problem of the core-halo mass relationship and a toy model for the TF regime of SFDM can be found in \citep{RS}. Those authors suggested that the wave nature of SFDM would result in an
effective pressure support (reflecting random wave motions and the associated density inhomogeneity implied on small scales) when averaged over scales larger than the core, leading to virial equilibrium on scales well beyond the size
of the core, out to the same scales as in CDM halos, in fact. Halo envelopes were later confirmed in the fuzzy regime through simulations by \citep{sc10} and follow-up studies, as referenced above.}  

{In the first part of this work, we will extend the core-halo mass relation \eqref{mcmh} to SFDM models with self-interaction, using analytical calculations. As for the fuzzy regime, the core mass increases with halo mass. In the second part, we will apply our result to assess the SFDM parameter space for which soliton cores eventually become too massive to remain stable: beyond a critical mass, which depends upon SFDM parameters, soliton cores will collapse and can form black holes.  In this way SFDM might provide a mechanism to form SMBHs in the centers of halos as of an early time. However, it turns out that it is very difficult to produce SMBHs with fiducial values of SFDM in the fuzzy regime, while SMBH formation is much more feasible, once self-interaction is added, as we will show.}

There is a host of observations that indicate the existence of SMBHs -- with masses ranging between $10^6-10^{10}~ M_{\odot}$, placed in the center of most massive galaxies \citep{sm2,sm3}. 
The origin of SMBHs is still mysterious, given their huge masses at the high redshifts ($z>5.6$), where they have been observed \citep{smbh1,smbh2,smbh3,smbh4,smbh5, smbh6,smbh7,smbh8,smbh9,smbh10,smbh11,smbh12,smbh13,smbh14}. 
{In order for stellar BHs to become supermassive, they would need to accrete large amounts of baryonic material and DM over a short time, which is unfeasible even if accretion happens at maximum Eddington rate.} In addition to this puzzle of high-$z$ SMBHs, there is also a problem in understanding why there seem to be no medium-sized black holes with masses $\sim 10^2-10^5 M_\odot$. 
Some standard scenarios of the formation of SMBHs consider the following: like stellar black holes (BH) which result from the collapse of massive stars, SMBHs could be produced either by the collapse of massive clouds of gas during the early stages of formation of a galaxy \citep{cloud_c}, or by the collapse of supermassive Pop III stars\footnote{These scenarios are not to be confused with another (non-standard) proposal to explain SMBHs, namely so-called supermassive ``dark stars", primordial stars of supermassive size which are powered by DM self-annihilation in models of WIMP and related dark matter, \citep{KF,R15}. Once dark stars collapse, they could form seed black holes of about $10^4-10^{5} M_\odot$.}. Another suggestion considers the formation of a cluster of stellar BHs, which eventually merge into a SMBH \citep{merger}. However, it seems that these scenarios do not deliver a fully satisfactory explanation for the formation and evolution of such SMBHs at high redshifts. Additionally, observations show that the masses $M_{SMBH}$ of the central SMBHs are correlated with various global properties of their host galaxies. The most important relationship concerns the mass of the SMBH and the bulge mass, and an even tighter correlation with the stellar velocity dispersion of the host galaxy bulge, first reported by \citep{correlation1} and \citep{correlation2}. As a result, it has been also suggested that the central SMBH mass is correlated with the total mass of its host galaxy \citep{smbhcorr,smbhcorr2}.

Observations might thus indicate that the formation and growth of SMBHs over time could be related to the DM-dominated galactic halos. With this in mind, \citep{sm7} studied the possibility that SMBHs might form by collapse of all or part of gravitationally-bound equilibrium objects made of SFDM, which are assumed to model nuclear galactic halos. Several earlier works have considered this scenario and studied its plausibility. Among them, it was demonstrated that self-gravitating objects comprised of free scalar field configurations with masses larger than $0.6m_{pl}^2/m$, where $m_{pl}$ is the Planck mass, are able to collapse and form a BH \citep{boson_stars1,sm9,sm10,sm11,sm12,sm15,sm16}. For the specific case of a mass $m\sim 10^{-22}~\rm{eV}/c^2$, such configurations have a critical mass of collapse of $\sim 10^{13}M_\odot$. On the other hand, simulations in spherical symmetry demonstrated that only a part of the scalar field collapses to form a BH, while the remaining scalar field continues to surround the resulting BH for a very long time (longer than the age of the Universe), and can play the role of the DM halo of the galaxy \citep{sm11,sm15,sm16}. It is important to mention that these studies conclude that most of the scalar field configuration collapses into a BH, leaving only a small scalar field remnant for the halo. However, these analyses have been performed in spherical symmetry and in a limited region of parameter space, corresponding to typical systems known as boson stars (BS). Also, in these studies, those BSs have been used to model the entire SFDM halo, an unrealistic scenario, because simulations by \citep{sc4,sc10,sc11,sc12,sc13,sc14,moczext1,moczext2} revealed that SFDM halos possess a more complicated core-envelope structure (see equation \eqref{mcmh}). However, the above results are nevertheless useful, given that such BSs represent very well the soliton profiles observed in the central region of a galactic halo and then, given the fact that we are interested in an extension of the model to include self-interaction, simulations for the case of a self-interacting BS should be also applicable to the central soliton in halos. This case is of interest for the issue of studying the possible collapse of the central soliton in the most massive galactic halos (hosting the most massive galaxies). In fact, this last scenario is one of the main objectives of our study that will be analysed in this work.

The paper is organised as follows: in section II we review the basic equations necessary to describe the self-interacting SFDM model: the Einstein-Klein-Gordon (EKG) system for a general description and the Gross-Pitaevskii-Poisson (GPP) system in the weak-field limit\footnote{In what follows, ``weak field" refers to the regime of weak gravitational fields, i.e. the Newtonian regime.}.
In section III, we present a basic description of the SFDM soliton profile, which is obtained as the minimum-energy, coherent, quasi-stationary solution of the GPP. In the same section, we consider a Gaussian ansatz to describe the soliton in order to maintain some freedom in working with the self-interaction parameter of the SFDM model. We show that, in general, this ansatz maintains practically all the relations that are found in the numerical description (the parameter dependence for the maximum mass for collapse of the soliton, parameter dependence in the TF regime, etc), even if relativistic corrections are considered. This implies that the Gaussian ansatz represents a good approximation for the soliton. {Also in this section, we show that the results provided by this ansatz can be easily obtained in the hydrodynamic representation of the GPP system and by considering a simple dimensional argument, without the need of considering any functional form for the core profile.} In section IV, we extend the core-halo mass relation to self-interacting SFDM by assuming that {some energy relations that }
are fulfilled by core and halo quantities in the {simplest} SFDM model {remain valid in the self-interacting scenario}. 
Because two relations have been reported between the masses of the core and the halo, we decided to extend both of them.
In section V, we compare our results with previous works, with the emphasis on the implied constraints of the SFDM model parameters. For this comparison, we focus on the core properties found in the central region of SFDM halos. We find that for a repulsive SFDM candidate, the central soliton remains stable and should be represented in the TF regime, while for attractive SFDM, we have scenarios where the soliton can collapse and form a SMBH in the most massive galacitc halos (hosting the most massive galaxies), while the cores remain stable in those halos that host the least massive galaxies.  Finally, in section VI we present our conclusions.

\section{\label{sec:system}Basic equations for the Scalar Field Dark Matter Model}

In this section, we review the basic equations necessary to describe the dynamics of a scalar field minimally coupled to gravity. We consider a complex field, given that the case of a real field is easily derived from this description. 

\subsection{The Einstein-Klein-Gordon system}

The set of differential equations governing the dynamics of a self-interacting scalar field minimally coupled to gravity is described by the EKG system: 
\begin{subequations}\label{EKGs}
\begin{equation}
\Box\varphi +2\frac{dV(|\varphi|^2)}{d|\varphi|^2}\varphi=0,
\end{equation}
\begin{equation}
R_{\alpha\beta}-\frac{1}{2}g_{\alpha\beta}R = \frac{8\pi G}{c^4}  T_{\alpha\beta},
\end{equation}
\end{subequations}
where 
\begin{equation}\label{potential}
 V(|\varphi|^2)=\frac{m^2c^2}{2\hbar^2}|\varphi|^2+\frac{\lambda}{4\hbar c}|\varphi|^4,
\end{equation}
$c$ is the speed of light, $\hbar$ is the reduced Planck constant, $G$ is the gravitational    constant, $\lambda$ is a self-interaction parameter that can be positive (repulsive) or negative (attractive), $\Box\equiv \nabla_\mu \nabla^\mu$ is the $4$-D'Alembert operator, $\varphi$ is the scalar field with units $[\sqrt{kg\cdot m}/s]$, $g_{\alpha\beta}$ is the spacetime metric, $R_{\alpha\beta}\ (R)$ is the Ricci tensor (scalar) and $T_{\alpha\beta}$ is the stress energy tensor which possesses all the energy components that exist in the system. Particularly, for the scalar field
\begin{eqnarray}
T_{\alpha\beta}^{(\varphi)} =&& \frac{1}{2}(\nabla_\alpha\varphi)^* (\nabla_\beta\varphi)+\frac{1}{2}(\nabla_\alpha\varphi) (\nabla_\beta\varphi)^* \nonumber\\
&&-g_{\alpha\beta}\left[\frac{1}{2}(\nabla^\gamma\varphi)^* (\nabla_\gamma\varphi)-V(|\varphi|^2)\right],
\end{eqnarray}
Here, Greek letters range from $0$ to $3$, denoting spacetime indices.

\subsection{The weak-field limit}

Structure formation of halos in a matter-dominated Universe can be well described within the weak-field regime. In this regime, \eqref{EKGs} is reduced to the Gross-Pitaevskii-Poisson (GPP) system \citep{wf1}
\begin{subequations}\label{schpo}
\begin{equation}
    i\hbar \frac{\partial \psi}{\partial t}=-\frac{\hbar^2}{2m}\nabla^2 \psi+m\Phi \psi+g|\psi|^2\psi, 
\end{equation}
\begin{equation}\label{Poisson_new}
\nabla^2\Phi = 4\pi G\rho,    
\end{equation}
\end{subequations}
where $\psi$ is defined in terms of $\varphi$ as
\begin{equation}\label{twoscalars}
    \varphi(\text{\textbf{x}},t) = \frac{\hbar}{\sqrt{m}}e^{-imc^2 t/\hbar}\psi(\text{\textbf{x}},t),
\end{equation}
$g\equiv \lambda\hbar^3/(2m^2 c)$, $\Phi$ is the gravitational potential and $\rho$ is a cosmological overdensity that usually possesses contributions from the DM and the baryonic components. If we ignore the baryonic contribution  (a limitation shared with most of the simulation work \citep{sc4,sc10,sc11,sc12,sc13,sc14}), we have $\rho = m|\psi|^2$. 

{Observe that by using in \eqref{potential} the new field $\psi$ defined in \eqref{twoscalars} and the definition of $g$, we obtain}
\begin{equation}
    {V(|\psi|^2) = \frac{mc^2}{2}|\psi|^2+\frac{g}{2}|\psi|^4},  
\end{equation}
{which is the scalar field potential in standard physical units.}

Two important quantities that are necessary to describe SFDM halos are the total mass $M_t$ and total energy $E_t$ associated with the system:
\begin{subequations}\label{mass_en}
\begin{equation}
    M_t = m\int_V |\psi|^2 d^3\text{\textbf{r}},
\end{equation}
\begin{equation}\label{Et}
    E_t = \int_V\left[\frac{\hbar^2}{2m}|\nabla\psi|^2+\frac{m}{2}\Phi|\psi|^2+\frac{g}{2}|\psi|^4\right]d^3\text{\textbf{r}}.
\end{equation}
\end{subequations}
Observe that the total energy can be written in a very instructive way
\begin{equation}\label{totalE}
    E_t = K_t + W_t + U_{SI,t},
\end{equation}
where 
\begin{subequations}\label{energies}
\begin{equation}\label{kt}
    K_t = \int_V\frac{\hbar^2}{2m}|\nabla\psi|^2d^3\text{\textbf{r}},  \end{equation}
is the total kinetic energy,
\begin{equation}\label{wt}
    W_t = \int_V\frac{m}{2}\Phi|\psi|^2d^3\text{\textbf{r}},
\end{equation}
is the total gravitational potential energy and 
\begin{equation}
    U_{SI,t}=\int_V \frac{g}{2}|\psi|^4d^3\text{\textbf{r}},
\end{equation}
is the total energy associated with the self-interaction. This last way of writing each energy contribution is very convenient, because they also appear in the scalar virial theorem of an isolated mass distribution 
\end{subequations}
\begin{equation}\label{virial}
    2K_t+W_t+3U_{SI,t}=0.
\end{equation}

On the other hand, notice that if we use the following variables
\begin{equation}
    \hat\psi = \sqrt{\frac{4\pi G \hbar^2}{mc^4}}\psi, \ \ \hat{\text{\textbf{r}}} = \frac{mc}{\hbar}\text{\textbf{r}},\ \ \ \hat t = \frac{mc^2}{\hbar}t\nonumber,
\end{equation}
    \begin{equation}\label{parameters}
    \hat \Phi = \frac{\Phi}{c^2}, \ \ \ \  \hat \Lambda = \frac{c^2g}{4\pi G\hbar^2}=\frac{m_{pl}^2}{m^2}\frac{\lambda}{8\pi},
\end{equation}
(note: $mc/\hbar$ is the inverse Compton length and $mc^2/\hbar$ is the bare angular frequency of the field)
the GPP system can be rewritten in a way where all the natural constants disappear:
\begin{subequations}\label{dimensionless}
\begin{equation}
    i\frac{\partial\hat\psi}{\partial \hat t} = -\frac{1}{2}\hat\nabla^2\hat\psi + \hat\Phi \hat\psi+ \hat\Lambda |\hat\psi|^2\hat\psi,
\end{equation}
\begin{equation}
    \hat\nabla^2\hat\Phi = |\hat\psi|^2.
\end{equation}
\end{subequations}
Additionally, there is a re-scaling property for this GPP system given by 
\begin{equation}\label{tilde}
    \lbrace \hat t,\hat{\text{\textbf{r}}},\hat \Lambda, \hat\psi,\hat\Phi\rbrace\ \ \ \Rightarrow\ \ \  \lbrace\gamma^{2}\hat t,\gamma\hat{\text{\textbf{r}}},\gamma^2 \hat \Lambda, \gamma^{-2}\hat\psi,\gamma^{-2}\hat\Phi\rbrace,
\end{equation}
where $\gamma > 0$ is a scaling parameter.
 Then, the different physical quantities defined in \eqref{mass_en} and \eqref{energies} are also re-scaled with similar relations.

\section{Soliton properties: general considerations}\label{sec:solitoncons}
\subsection{The weak-field limit}

It is now  accepted that at large time scales (structure formation time scales) the averaged density profile of cores appearing in central regions of SFDM haloes can be well fitted by coherent, quasi-stationary, ground-state solutions of the GPP system. In this section, we review different previous results which we will use in order to extend the soliton description to self-interacting SFDM particles. 

Quasi-stationary states of \eqref{dimensionless} are described by dimensionless wavefunctions of the form
\begin{equation}\label{ansatz}
\hat\psi(\hat{\text{\textbf{r}}},\hat t)=\hat\phi(\hat{{r}})e^{-i\hat\mu \hat t}, \ \ \ \hat\mu,\hat\phi\in \mathbb{R},
\end{equation}
where $\hat r$ is the dimensionless spherical radial coordinate and $\hat\mu$, the dimensionless GPP chemical potential, should be fixed by the conservation of particle number. In general, the system \eqref{dimensionless} with the ansatz \eqref{ansatz} has an infinite number of different solutions that fulfill appropriate initial and boundary conditions \citep{ure-guz,wf3}\footnote{The typical boundary conditions are given by regularity in the origin $\hat\phi(\hat r=0)=\hat\phi_0$, $\hat\phi'(\hat r=0)=0$ and asymptotic vanishing $\hat\phi(\hat r\rightarrow\infty)\rightarrow 0$, $\hat\Phi(\hat r\rightarrow \infty)\simeq -M/r^2$. 
}. Each of them -- usually called Newtonian boson stars (NBS) -- can be identified by the number of nodes of $\hat\phi$, before the solution decays asymptotically. The solution without nodes -- the soliton -- is considered the ground state of the GPP system and it possesses the smallest value of $\hat\mu$, while solutions with nodes are usually considered as excited NBSs.    

Observe that from \eqref{tilde} it is possible to construct different solutions for the soliton, once one of them is known. In fact, as explained in \citep{wf3}, in the free case it is possible to construct all the ground state solutions for a given central scalar field value, just by using the re-scaling property in \eqref{tilde}. On the other hand, in the self-interacting case something similar occurs: once a ground state solution is known for a given value of $\hat\Lambda$, it is possible to construct all the ground state solutions for different central value of the scalar field and the same value of $\hat\Lambda$, just by using the re-scaling properties provided in \eqref{tilde}. However, if we were interested in finding a new soliton with a different $\hat\Lambda$, it would be necessary to solve the differential equations \eqref{dimensionless} for such $\hat \Lambda$ again. Therefore, we can see that once a self-interaction parameter is added to the model, we do not have the same freedom in working with the soliton solution, as in the free case. Nevertheless, as we shall see in this section, we can avoid this problem, once a Gaussian approximation is adopted. {In fact, in this section we shall also show that the results obtained from the Gaussian ansatz can be reconstructed by considering a dimensional argument.}  To this end, let us continue to present some basic relations that will be helpful for our later description, and apply them to the case of the free field in order to compare with our Gaussian ansatz later.

First of all, all soliton solutions are virialized structures that fulfill equation \eqref{virial}. 

Now, let us focus on the free case: Usually one solves for that solution for which the central value $\hat \psi(\hat r=0)=1$. In this case, the numerical value of the dimensionless chemical potential is $\hat\mu \simeq -0.69$.
Such solution can be used together with the re-scaling parameter $\gamma$ in \eqref{tilde} to construct solitons with different masses by fixing the $\gamma$ parameter as \citep{wf2}: 
\begin{equation}
    \gamma = 3.6\times10^{-6}m_{22}M_{c,7}^{(\gamma)},
\end{equation}
where $M_{c,7}^{(\gamma)}\equiv M_c^{(\gamma)}/(10^7 M_\odot)$ and 
\begin{displaymath}
m_{22}\equiv m/(10^{-22}eV/c^2).
\end{displaymath}
Notice that we have left explicitly the $\gamma$ dependence for the numerical solution of the soliton.

Another important quantity is the radius that contains $99\%$ of the soliton mass,
\begin{equation}\label{RMrelation}
 R_{99}^{(\gamma)} =9.9\frac{\hbar^2}{GM_c^{(\gamma)}m^2},
\end{equation}
(see for example \citep{sf12}),
or in fiducial notation
\begin{equation}\label{R_99}
    R_{99}^{(\gamma)}\simeq \frac{8.445\times 10^{4}}{(m_{22})^2M_{c,7}^{(\gamma)}} \ \rm{pc}.
\end{equation}
Finally, the soliton fulfills the relations  
\begin{subequations}
\begin{equation}\label{M_c_E}
    M_c^{(\gamma)} \simeq 4.3\sqrt{\frac{|E_c^{(\gamma)|}}{M_c^{(\gamma)}}}\frac{m_{pl}^2}{mc},
\end{equation}
{
\begin{equation}\label{energy_mocz1}
    M_c^{(\gamma)}\simeq 2.6\left(\frac{|E_c^{(\gamma)}|}{(mG/\hbar)^2}\right)^{1/3},
\end{equation}
}
\end{subequations}
which was found by \citep{wf2}. 
In the free case, it is equivalent to
\begin{subequations}
\begin{equation}\label{M_c_K}
    M_c^{(\gamma)} \simeq 4.3\sqrt{\frac{K_c^{(\gamma)}}{M_c^{(\gamma)}}}\frac{m_{pl}^2}{mc} = 4.3\sqrt{\frac{|W_c^{(\gamma)}|}{2M_c^{(\gamma)}}}\frac{m_{pl}^2}{mc},
\end{equation}
{ 
\begin{equation}\label{energy_mocz2}
    M_c^{(\gamma)}\simeq 2.6\left(\frac{K_c^{(\gamma)}}{(mG/\hbar)^2}\right)^{1/3}=2.6\left(\frac{|W_c^{(\gamma)}|}{2(mG/\hbar)^2}\right)^{1/3},
\end{equation}
}
\end{subequations}
which has been pointed out in \citep{core_halo1}.

\textit{Remark:} We note that thanks to the re-scaling property in the GPP system based upon the Newtonian approximation, the configuration does not admit an upper critical mass. However, from general-relativistic calculations follows a limiting maximum mass beyond which collapse to a BH occurs. On the other hand, if $\hat\Lambda<0$, there is a maximum mass, even for a Newtonian soliton, given by \citep{boson_stars}
\begin{equation}\label{Mln}
    M_{c,max}\simeq 10.03\frac{m_{pl}}{\sqrt{|\lambda|}},
    \end{equation}
where we have decided to use $\lambda$ instead of $\hat\Lambda$ for simplicity in the  expression; the $\hat\Lambda$ dependence for the above critical mass can be easily obtained from \eqref{parameters}.

\subsubsection{The Gaussian ansatz in the weak-field limit}\label{gaussian}

Previous literature has made extensive use of two different analytic approximations for the numerical, exact soliton profile of SFDM halos without self-interaction. On the one hand, there is a rational function approximation, which was proposed in \citep{sc4}, and which is based upon an empirical fit to the central region of simulated halos. On the other hand, a Gaussian profile has been used to approximate SFDM solitons in \citep{sc15,guzman2018head}.
The use of a Gaussian is motivated by the fact that Gaussian ``wave packets" not only appear in many contexts of a linear Schr\"odinger equation, it also constitutes a solution for laboratory Bose-Einstein condensates without particle self-interaction,  see e.g. \citep{BP}. In this work, we decided to use the Gaussian approach, given its better physical foundation and the fact that it is easier to find physical relations of interest from it, given the difficulties described above with respect to the quasi-stationary states of variable $\hat \Lambda$. The difference between the two analytic
profiles can be seen in appendix \ref{ap:gvsp}: the rational function description appears to match better the numerical result for the soliton if $\hat r$ is small,
while the Gaussian 
approach matches better the numerical solution at large $\hat r$. However, what is more important in our context is the fact that the Gaussian ansatz arrives at the same physical relationships than the numerical solution, only the prefactors differ by factors of a few.

\begin{figure}
    \centering
    \includegraphics[width=3.4in]{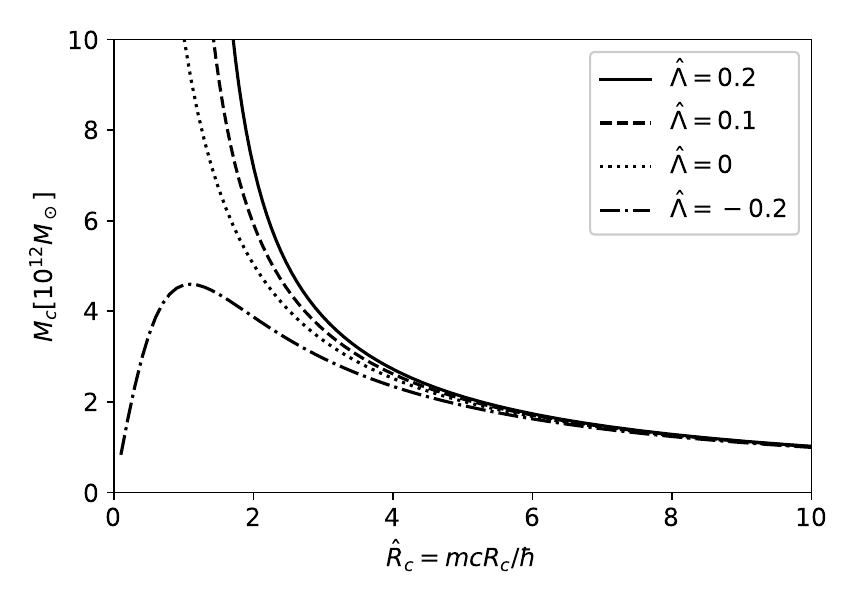}
    \caption{\footnotesize{Mass-radius relation of the soliton in self-interacting SFDM models: we plot equ.(\ref{m_r_self2}) for $m_{22}=1$ and various self-interaction strengths $\hat \Lambda$.}}
    \label{mass_radius}
\end{figure}

{Now, the question arises to what extent the Gaussian can be used, if self-interaction is included.  In fact, \citep{BP} already used a Gaussian ansatz as a trial function in a variational analysis, in order to find modified physical relationships, valid when self-interaction is included. The same approach was proposed in
\citep{sc15} in order to extend the modeling of the SFDM soliton profile with self-interaction by considering the Gaussian density distribution}
\begin{equation}\label{deng1}
    \rho_c^{(g)}(r)= \frac{M_c}{(\pi R_c^2)^{3/2}}e^{-r^2/R_c^2},
\end{equation}
where $R_c$ is a characteristic core radius associated with the radius that contains $99\%$ of the total mass of the distribution\footnote{{This number follows simply by calculating the radius which includes $99\%$, i.e. $2\sigma$ of the mass of the Gaussian distribution.}} as $R_{99}=2.38167 R_c$. 
For the sake of the reader, let us quote some of the results which follow from this approach:
A mass-radius relation was found by way of minimizing the energy\footnote{{Notice that \citep{sc15} uses an uncommon definition of $W$ which differs from equ.\eqref{wt} by a factor of $1/2$.}} functional \eqref{Et} or \eqref{totalE} and considering the ansatz \eqref{deng1} as a trial function. 
That procedure yields
\begin{equation}\label{m_r_self1}
    M_c = 3\sqrt{{2\pi}}\frac{\frac{\hbar^2}{Gm^2 R_c }}{1-\frac{6g}{4\pi Gm^2 R_c^2}},
    \end{equation}
or equivalently,
\begin{equation}\label{m_r_self2}
    M_{c,7}\simeq \frac{10.076\times 10^{5}}{m_{22}}\frac{\hat R_c}{\hat R_c^2-6\hat\Lambda},
\end{equation}
 which is plotted in figure \ref{mass_radius} .
We can solve this equation for the radius, in turn,
\begin{equation}\label{radius_core}
    R_c = \frac{3\sqrt{2\pi}\hbar^2}{2 G m^2 M_c}+\sqrt{\left(\frac{3\sqrt{2\pi}\hbar^2}{2 G m^2 M_c}\right)^2+ \left(\frac{6g}{4\pi Gm^2 R_c^2}\right)^2},
\end{equation}
or in fiducial notationº:
\begin{equation}\label{R_cs}
    \hat R_c =  \frac{5.04\times 10^5}{M_{c,7}m_{22}}\left[1+\sqrt{1+6\hat\Lambda \left(\frac{M_{c,7}m_{22}}{5.04\times 10^5}\right)^2}\right].
\end{equation}
For $\hat \Lambda = 0$, we recover the mass-radius relationship in the free case.
On the other hand, when $\hat\Lambda>0$ and if the second term in the square root dominates, we obtain $\hat R_c\simeq \hat R_{c}^{(TF)}\equiv \sqrt{6\hat \Lambda}$, 
independent of $M_{c,7}$. In physical units, the radius in this regime of strong self-interaction reads $R_c \simeq \sqrt{6g/(4\pi G m^2)}$, which recovers the form of the so-called Thomas-Fermi radius
\begin{equation} \label{TFradius}
R^{(TF)}=\pi\sqrt{\frac{g}{4\pi Gm^2}},
\end{equation}
which corresponds to the radius of an $(n=1)$-polytrope (see \citep{goodman2000repulsive,peebles2000fluid})\footnote{
We can compare the radius that contains $99\%$ of the total mass of the Gaussian ansatz and the radius obtained in the TF regime: Considering that for the Gaussian ansatz \citep{sc15} $R_{99}\simeq 2.38167R_c^{(TF)} \simeq 5.834\sqrt{g/(4\pi Gm^2)}$, we can see that both quantities are close within a factor of $2-3$. }.

 

By comparing figure \ref{mass_radius} and the numerical solution (figure 1 in \citep{wf3}), we can see that the relation \eqref{m_r_self1} or \eqref{m_r_self2} maintains the same basic parameter dependence than the numerical solution. For the attractive case ($\hat\Lambda<0$), this means that there exists a maximum mass allowed by the scalar field configuration given by
\begin{equation}\label{maximum_mass_neg}
M_{c,max} \simeq 7.70 \frac{m_{pl}}{\sqrt{|\lambda|}}, 
\end{equation}
(see also equation (\ref{Mln})),
while in the repulsive and free case ($\hat\Lambda\geq 0$), there is no maximum mass for the soliton (the mass is unbound). Another important property that is also maintained is the fact that, the larger the coefficient $\hat\Lambda$, the more massive the equilibrium configuration. Interestingly, the radius at which the attractive case reaches its maximum mass is given by $\hat R_c= \hat R_{c,min} \equiv \sqrt{6|\hat\Lambda|}$, which\footnote{Notice that in refering to this radius we use the subscript ``min". The reason we use this subscript is because, as we will see later, this critical mass also corresponds to the minimum radius at which these configurations remain stable (see also \citep{sc15}).} is just the same radius at which the repulsive case goes over to the TF regime.

Now, we can easily show that the different energies defined in \eqref{totalE} and \eqref{energies} are as follows for the SFDM Gaussian ansatz with self-interaction: 
\begin{subequations}\label{energies_gaussian}
{
\begin{equation}\label{energy_c}
    {E_c} = -\left[\frac{3\hbar^2M_c}{4m^2R_c^2}+\frac{gM_c}{2\sqrt{2}\pi^{3/2}m^2 R_c^3}\right],
\end{equation}
\begin{equation}\label{kcg}
    K_c = \frac{3}{4}\frac{\hbar^2 M_c}{m^2 R_c^2},
\end{equation}
\begin{equation}
    W_c = -\frac{GM_c^2}{2\sqrt{2\pi}R_c},
\end{equation}
\begin{equation}
    U_{SI,c} = \frac{gM_c}{4\pi \sqrt{2\pi}m^2 R_c^3},
\end{equation}
}
\end{subequations}
{where in the expression for $E_c$ we have used the virial theorem (\ref{virial}), applied to the core.} {Equation \eqref{kcg}} can be re-arranged to
\begin{equation}
    \frac{1}{R_c} = \frac{2m}{\sqrt{3}\hbar}\left(\frac{K_c}{M_c}\right)^{1/2}.
\end{equation}
Using the $M_c - R_c$ relation (see equation \eqref{m_r_self1}), we arrive at
\begin{equation}\label{mc_km}
    M_c \simeq 8.68\frac{m_{pl}^2}{m} \frac{\left(\frac{K_c}{M_c}\right)^{1/2}/c}{1-8\hat\Lambda\left(\frac{K_c}{M_c}\right)/c^2}.
\end{equation}
Observe that in the free case ($\hat\Lambda = 0$) the above result differs from the result of simulations in eqn. \eqref{M_c_K} only by a factor of 2. {In a similar way, by interchanging $K_c\rightarrow -E_c$ (by using the virial theorem in the free-field limit), the above expression differs from \eqref{M_c_E} by a factor of 2 as well.}

{We can re-express the energies in \eqref{energies_gaussian} per core mass (i.e. specific energies), using the mass-radius relation \eqref{m_r_self1} as}
\begin{subequations}\label{energies_ek}
\begin{equation}\label{Energy_em}
    \frac{E_c}{M_c} = -\frac{1}{4\sqrt{2\pi}}\left(\frac{GM_c}{R_c}\right)\left[1\pm \frac{1}{3}\left(\frac{R_{crit}}{R_c}\right)^2\right],
\end{equation}
\begin{equation}\label{kinetic_em}
    \frac{K_c}{M_c} = \frac{1}{4\sqrt{2\pi}}\left(\frac{GM_c}{R_c}\right)\left[
1\mp\left(\frac{R_{crit}}{R_c}\right)^2\right],
\end{equation}
\begin{equation}\label{gravitational_em}
    \frac{W_c}{M_c} = -\frac{1}{4\sqrt{2\pi}}\left(\frac{GM_c}{R_c}\right)\left[2\mp \frac{3}{2}\left(\frac{R_{crit}}{R_c}\right)^2\right],
\end{equation}
\begin{equation}
    \frac{U_{SI,c}}{M_c} = \frac{1}{4\sqrt{2\pi}}\left(\frac{GM_c}{R_c}\right)\left[\pm\frac{1}{6}\left(\frac{R_{crit}}{R_c}\right)^2\right],
\end{equation}
\end{subequations}
where the upper (lower) sign is for a repulsive (attractive) self-interaction, and in the above expressions we have defined $R_{crit} \equiv \sqrt{6|g|/(4\pi G m^2)}$, in such a way that when $g>0$, $R_{crit} = R_c^{(TF)}$, whereas when $g<0$, $R_{crit} = R_{c,min}$.

{Finally, observe that in} the free case, we obtain from \eqref{R_cs},  expressed in fiducial units 
\begin{equation}\label{R_c}
R_c \simeq \frac{6.44\times 10^4}{(m_{22})^2 M_{c,7}} \ \rm{pc}.  
\end{equation}
The radius $R_{99}$ that contains $99\%$ of the total mass of the soliton is $R_{99}= 2.38167 R_c$, which from the above equation yields
\begin{equation}\label{core_halo_gaussian}
R_{99} \simeq \frac{15.34\times 10^4 }{(m_{22})^2 M_{c,7}}\ \rm{pc},
\end{equation}
as compared to the numerical result in \eqref{R_99}. 

{Additionally, if we use equations \eqref{Energy_em}, \eqref{kinetic_em} and \eqref{gravitational_em} together with our mass-radius relation \eqref{m_r_self1}, we can express the core mass in the free-field limit as
\begin{subequations}\label{ec_gauss}
\begin{equation}\label{energymocz1}
    M_c\simeq 4.22\left(\frac{|E_c|}{(mG/\hbar)^2}\right)^{1/3}.
\end{equation}
\begin{equation}\label{energymocz2}
    M_c\simeq 4.22\left(\frac{K_c}{(mG/\hbar)^2}\right)^{1/3}.
\end{equation}
\begin{equation}\label{energymocz3}
    M_c\simeq 4.22\left(\frac{|W_c|}{2(mG/\hbar)^2}\right)^{1/3}.
\end{equation}
\end{subequations}}
Comparing {\eqref{core_halo_gaussian} with} the numerical result in \eqref{R_99}{, or \eqref{ec_gauss} with \eqref{energy_mocz1} and \eqref{energy_mocz2}}, we see that the difference {between the results from the Gaussian ansatz versus the exact numerical solution are} small, of factors of a few. 

\subsubsection{Understanding the mass-radius relation for the solitonic core from the hydrodynamic representation of the GPP system}\label{heuristic}

Although the Gaussian ansatz has been extensively used in the literature and is well motivated to represent the numerical ground state solution of the GPP system, we include another derivation in this subsection, independent of the functional form of the ``trial function". For that purpose, we use the hydrodynamic representation of the GPP system. 

By decomposing the wavefunction $\psi$ in polar form as
\begin{equation}\label{densityrho}
    \psi(\text{\textbf{r}},t) = \sqrt{\frac{\rho(\text{\textbf{r}},t)}{m}}e^{iS(\text{\textbf{r}},t)},
\end{equation}
and defining a velocity field as
\begin{equation}\label{velocityv}
    \bar v = \frac{\hbar}{m}\nabla S,
\end{equation}
the GPP system is rewritten as an Euler and a continuity equation given by
\begin{subequations}\label{hydroeq}
\begin{equation}
    \rho\frac{\partial \bar v}{\partial t}+\rho(\bar v\cdot \nabla)\bar v = -\rho \nabla Q-\rho\nabla \Phi-\nabla P_{SI},
\end{equation}
\begin{equation}\label{continuity}
    \frac{\partial \rho}{\partial t}+\nabla\cdot(\rho \bar v) = 0,
\end{equation}
\end{subequations}
where
\begin{equation}
    Q \equiv -\frac{\hbar^2}{2m^2}\frac{\nabla^2\sqrt{\rho}}{\sqrt{\rho}} \ \ \ \ \text{and} \ \ \ \ P_{SI}\equiv \frac{g}{2m^2}\rho^2.
\end{equation}
    The term $Q$ is known as the quantum potential which arises from the quantum nature of SFDM, while $P_{SI}$ can be interpreted as a pressure term that is generated by the self-interaction between SFDM particles. In order to understand the parameter dependence of the soliton profile in the self-interacting SFDM model, we consider the following simplification: soliton structures fulfil $\partial \bar v/\partial t = 0 = \bar v$. Also, for simplicity we set $\nabla \sim 1/\mathcal{R}_c$, where $\mathcal{R}_c$ is the characteristic radius of the system, then
\begin{equation}
    \nabla Q\sim -\frac{\hbar^2}{2m^2\mathcal{R}_c^3}, \ \ \ \ \nabla P_{SI}\sim -\frac{g\rho^2}{2m^2 \mathcal{R}_c}\nonumber,
    \end{equation}
    \begin{equation}
    \nabla\Phi\sim \frac{GM_c}{\mathcal{R}_c^2},
\end{equation}
{where we choose the sign in $\nabla P_{SI}$ in such a way that this term correctly describes an attractive/repulsive pressure term, which is also consistent with the Gaussian ansatz.} Using $\rho \sim 3M_c/(4\pi \mathcal{R}_c^3)$, which is equivalent to saying that the soliton profile possesses a nearly constant density, we obtain from \eqref{hydroeq}
\begin{subequations}
\begin{equation}\label{v2}
    -\frac{\hbar^2}{2m^2 \mathcal{R}_c^2}+a\frac{GM_c}{\mathcal{R}_c}-b\frac{3g M_c}{8\pi m^2 \mathcal{R}_c^3} = 0,
\end{equation}
\begin{equation}\label{rhot}
    \frac{\partial \rho}{\partial t} = 0,
\end{equation}
\end{subequations}
where $a$ and $b$ are some constants that we introduced to apply the summation in the above expression (i.e. we are considering, for example, that $\nabla Q\simeq const*\hbar^2/(2m^2\mathcal{R}_c^3)$). 

First, equation \eqref{rhot} reflects our assumption of a stationary solution, which is in agreement with \eqref{ansatz}. On the other hand, from \eqref{v2} and considering that $g>0$ we obtain
\begin{equation}\label{equilibriumpos}
    a\frac{GM_c}{\mathcal{R}_c}= \frac{\hbar^2}{2m^2 \mathcal{R}_c^2} +b\frac{3g M_c}{8\pi m^2 \mathcal{R}_c^3},
\end{equation}
and so it is easy to see that solitons are produced by the equilibrium between self-gravity (left-hand side in the above expression) and the pressures due to quantum uncertainty and self-interaction.

Two well studied limit cases are
\begin{itemize}
    \item \textbf{The fuzzy limit:} This regime is obtained when the second term on the right-hand side of equation \eqref{equilibriumpos} can be ignored, and then the soliton is a result of the equilibrium between quantum pressure and gravity. In this limit, the $M_c-\mathcal{R}_c$ relation reads 
    \begin{equation}
        M_c \mathcal{R}_c = \frac{1}{2a}\frac{\hbar^2}{Gm^2},
    \end{equation}
    which maintains the same parameter dependence found in the numerical treatment, see equation (\ref{RMrelation}).
    \item \textbf{The Thomas-Fermi approximation:} This regime is obtained when the first term on the right-hand side of equation \eqref{equilibriumpos} can be ignored, and then the soliton results as an equilibrium between gravity and the pressure due to self-interaction. In this limit, the soliton fulfills the $M_c-\mathcal{R}_c$ relation
    \begin{equation}
        \mathcal{R}_c=\sqrt{\frac{b}{2a}\frac{3g}{4\pi Gm^2}},
    \end{equation}
    which also maintains the same parameter dependence found by the exact solution, see equation (\ref{TFradius}).
\end{itemize}
On the other hand, if $g<0$ we have
\begin{equation}\label{repulsiveself}
    a\frac{GM_c}{\mathcal{R}_c}+b\frac{3|g| M_c}{8\pi m^2 \mathcal{R}_c^3}= \frac{\hbar^2}{2m^2 \mathcal{R}_c^2},
\end{equation}
and then the soliton results as an equilibrium between gravity plus self-interaction pressure and the repulsion due to the quantum pressure. Observe that in this scenario, we can also define a new limiting case
\begin{itemize}
    \item \textbf{The strong self-interaction regime in the attractive scenario:} This regime is obtained when the first term on the left-hand side of equation \eqref{repulsiveself} can be ignored, and then the soliton can be understood as the equilibrium between quantum pressure and attractive self-interaction. In this limit, the $M_c-\mathcal{R}_c$ relation is
    \begin{equation}
        \mathcal{R}_c = b\frac{3|g|}{4\pi \hbar^2}M_c.
    \end{equation}
These configurations correspond to soliton profiles with radius smaller than the one with the maximum possible mass shown in figure \ref{mass_radius}. However, as already mentioned in footnote 7, these configurations turn out to be unstable.
\end{itemize}

Re-arranging equation \eqref{v2} we have
\begin{equation}\label{mcrc2}
    M_c= \frac{1}{2a}\frac{\frac{\hbar^2}{Gm^2\mathcal{R}_c}}{1-\frac{b}{2a}\frac{3g}{4\pi Gm^2 \mathcal{R}_c^2}},
\end{equation}
and it is easy to see that this relation is equivalent to the one shown in \eqref{m_r_self1} from the Gaussian ansatz.

At this point, we have not yet specified the numerical values of $a$ and $b$. In order to do so, we could proceed in two different ways: First, we use the result from the Gaussian ansatz and set $\mathcal{R}_c =R_c$. In this case,  
\begin{equation}
    \frac{1}{2a} = 3\sqrt{{2\pi}}, \ \ \ \ \frac{b}{2a} = 2,
\end{equation}
and the $M_c-R_c$ relation is then given exactly by \eqref{m_r_self1}. On the other hand, we could also fix the numerical values of $a$ and $b$ by matching our result with the exact numerical solution. For example, let us suppose that $R$ is the radius that contains $99\%$ of the total mass of the configuration and that such a radius can always be written as $R = const*\mathcal{R}_c$. Then, from \eqref{mcrc2} we obtain
\begin{equation}
    M_c= \frac{1}{2\hat a}\frac{\frac{\hbar^2}{Gm^2R}}{1-\frac{\hat b}{2\hat a}\frac{3g}{4\pi Gm^2 R^2}},
\end{equation}
where $\hat a$ and $\hat b$ are new constants. By matching the last expression with the result in the free case (\ref{RMrelation}) and the TF regime (\ref{TFradius}), respectively, we have
\begin{equation}
    \frac{1}{2\hat a} = 9.9, \ \ \ \ \sqrt{\frac{\hat b}{2\hat a}} = \pi,
\end{equation}
and the final $M_c-R$ relation should read
\begin{equation}\label{mcrc99}
        M_c= 9.9\frac{\frac{\hbar^2}{Gm^2R}}{1-\pi^2\frac{3g}{4\pi Gm^2 R^2}}.
\end{equation}

{We stress that this way of obtaining the mass-radius relations for the soliton is particularly interesting, because the only thing we needed to do was to consider the characteristic scales of the system. In all cases, this simple analysis reproduces correctly the main features already known from the numerical and analytical descriptions of the soliton. The only differences which occur involve the numerical values of the constants that accompany the parameter dependence of the different relations, and they are all within factors of a few.  
Nevertheless, for the sake of concreteness, we decided to continue to use the results obtained from the Gaussian ansatz for the rest of this work.  }

\subsection{Implications from a general-relativistic treatment}\label{relat_corr}

The analysis of the previous subsections was carried out in the weak-field regime. 
This regime is a good approximation, given that it is very well justified at galactic scales. Yet, it leaves out an important physical phenomenon, namely the fact that a limiting maximum mass is predicted to exist for the soliton, once general-relativistic effects are considered.
 
It is then natural to anticipate that for certain masses of the soliton,  a relativistic treatment should be important -- as it turns out, it is possible that some of the cores of SFDM halos are not covered by the weak-field limit, for example the cores of the most massive galaxies that possess the most massive solitons (see equation \eqref{mcmh} or the next section for the generalization to the self-interacting case).
In this circumstance, the correct way to model such solitons should be in the general relativistic scenario, i.e. by solving the EKG system \eqref{EKGs}. For this reason, in this subsection we review an important consequence obtained when relativistic effects are taken into consideration: the maximum mass beyond which the soliton will collapse to form a BH.

By analogy to the weak-field limit, we assume that the core profiles in the central region of galactic halos made of SFDM are given by the minimum-energy, coherent, quasi-stationary solution of the EKG system \eqref{EKGs}. We can obtain such solutions by demanding spherical symmetry. In this case, the spacetime for the self-gravitating scalar field can be well described by the metric
\begin{equation}\label{metric}
    ds^2 = -\alpha(r)^2dt^2+a(r)^2 dr^2+r^2d\Omega^2,
\end{equation}
where $\alpha$ and $a$ are real metric functions, $r$ is usually called areal radius and $d\Omega^2\equiv d\theta^2+\sin^2 \theta d\phi^2$ is the solid angle square differential. 
As it was shown in \citep{boson_stars1}, after considering a standard post-Newtonian treatment, the EKG system \eqref{EKGs}, with a geometry defined by the above metric, is reduced to the GPP equations \eqref{dimensionless} in the weak-field limit. 

The set of equations \eqref{EKGs} together with the metric \eqref{metric} have been extensively studied in the literature in the context of boson stars (BS). The way to construct such solutions is similar than in the Newtonian case (see \citep{dynamical} for a review and references therein), i.e. a harmonic time dependence for the scalar field is proposed, a central scalar field value $\varphi(0)$ is specified and the same kind of boundary conditions for the soliton solution than in the weak-field limit are imposed. In doing so, the final configurations that are obtained can be split into two regions -- a stable\footnote{The soliton profiles that are valid in the weak-field limit are part of this branch. Although the Newtonian approximation makes it appear as if it is possible to construct solitons with unlimited mass, it is important to realize that, within the Newtonian description, there exists a critical value of the parameter $\gamma$ beyond which it is not possible to construct solitons from the scaling property
\eqref{tilde}.} and an unstable branch -- divided by a maximum mass $M_{c,max}$ allowed by a BS made of scalar field\footnote{The way to know if a configuration will have a given dynamics proceeds by calculating the binding energy of the BS as
$E_b = M_{MS}(r\rightarrow \infty)-Nm$, where $M_{MS}$ is the mass of the BS enclosed within a given $r$, defined in terms of the Misner-Sharp mass function $M_{MS}= \frac{r}{2}\left(1-\frac{1}{a^2(r)}\right)$,
and $N$ is the total number of bosons. It happens that when $E_b>0$, the system has an excess of energy and will disperse. On the other hand, if $E_b<0$, the system is gravitationally bound and will collapse to a BH, if it is in the unstable branch. Otherwise, it will remain coherent, if it is in the stable branch.}. The stable branch is at higher radii (i.e. right side of maximum mass), while the unstable branch is at the left side of the maximum mass at smaller radii.
For masses bigger than $M_{c,max}$, stable BSs do not exist, and in such a case, configurations with masses $M>M_{c,max}$ should collapse to form a BH. The parameter dependence of the maximum mass in the free and repulsive self-interacting scenarios are as follows:
\begin{itemize}
    \item In the free case \citep{boson_stars1,sm9,sm10,sm11,sm12,sm15,sm16}:
\begin{equation}\label{mmasfree}
    M_{c,max}\simeq 0.633 \frac{m_{pl}^2}{m}.
\end{equation}
\item In the case of a scalar field with repulsive interaction, the maximum mass for stable configurations is given by \citep{colpi}
\begin{equation}\label{mass_c_pos}
    M_{c,max}\simeq  0.22\sqrt{\hat \Lambda}\frac{m_{pl}^2}{m}.
\end{equation}
\end{itemize}
Let us compare these results with ours from our ansatz. We may assume that the Gaussian will collapse into a BH, once $R_{99}=2.38167 R_c\simeq R_{sch}$, where $R_{sch}\equiv 2M_c G/c^2$ is the Schwarzschild radius associated with the soliton. By considering \eqref{R_cs}, expressing hat quantities in terms of physical ones with \eqref{parameters} and equating $R_{99}=R_{sch}$, we obtain in the free case $M_{c,max}\simeq 2.11 m_{pl}^2/m$, whereas in the strong, repulsive self-interaction regime $M_{c,max}\simeq 3.57\sqrt{\hat\Lambda}m_{pl}^2/m$. Note that in both cases the same parameter dependence is maintained for $M_{c,max}$ as for the general-relativistic results, with the only difference again in the numerical prefactors that accompany each relation. Of course, the difference between these prefactors is rooted in the fact that we are trying to match a Newtonian ansatz with a general-relativistic result and, as expected, the critical masses from general relativity are lower than the Newtonian analysis suggests.

\section{Core-halo structure in the self-interacting scenario}





The numerical simulations performed by several authors \citep{sc4,sc10,sc11,sc12,sc13,sc14,moczext1,moczext2} have revealed that halos made of SFDM without self-interaction show a core-envelope structure, where a central core transitions at a certain radius to an ``NFW-like" halo envelope.

Several attempts have been made to understand whether there are global relationships that allow the quantities of these central solitons to be related to properties of the halo. However, the correct way in which they are related is not yet fully understood, as several of these works have reported different functional relations between the masses of these cores and the total halo. For example, in \citep{sc4,sc10} it was reported from cosmological simulations a core-halo mass relation which we can write in a fiducial way as
\begin{equation}\label{M_c}
    M_{c,7} \simeq 1.4\times 10^2 \frac{M_{h,12}^{1/3}}{ m_{22}},
\end{equation}
where 
\begin{displaymath}
M_{h,12}\equiv M_h/(10^{12}M_\odot),
\end{displaymath}  and the subindex $h$ refers to halo quantities. Schive et al. also showed that in all the galaxies that they simulated, the final structures also fulfilled the energy relation
\begin{equation}\label{M_c_Eh}
    M_c \simeq 4.3\sqrt{\frac{|E_h|}{M_h}}\frac{m_{pl}^2}{mc}.
\end{equation}
Several authors (see for example \citep{chavanis2019core, chavanis2019predictive,sc13}) have reasoned that the above core-halo mass relationship could be explained, if the characteristic circular velocity at the core radius is roughly the same order than that at the halo radius (``velocity dispersion tracing"), i.e.
that the condition
\begin{equation}\label{vcvh}
    v_c\sim v_h \ \ \ \Rightarrow \ \ \ \ \frac{GM_c}{R_c}\sim \frac{GM_h}{R_h},
\end{equation}
should be fulfilled. The physical meaning of this relation is that the size of the soliton matches the de Broglie wavelength, {expressed with the velocity dispersion $\sigma$ of the halo}, resulting in a nontrivial type of non-local uncertainty principle; it has been also suggested that this relation follows from an equilibrium between the virial temperature of the core and the halo. On the other hand, {in \citep{wf2} it was suggested that, for an isolated soliton whose mass is written as equation \eqref{M_c_E}, and comparing with the result of Schive et al. \eqref{M_c_Eh}, the core-halo mass relation could be understood, if the specific energy for the central soliton and for the host halo are the same, i.e. if the condition 
\begin{equation}\label{ec_eh}
    \frac{|E_c|}{M_c}\simeq \frac{|E_h|}{M_h},
\end{equation}
applies. Observe that, from the virial theorem for a free SFDM candidate, the above condition also implies that
}
%
\begin{equation}\label{core_halo}
    \frac{K_c}{M_c}\simeq \frac{K_h}{M_h}.
\end{equation}
In a more recent work, in \citep{core_halo1} it was suggested that the latter relation was better suited to reproduce core-halo relations. However, so far, all of these three relations are being used in the literature to explain the physical nature of \eqref{M_c}. This is because, in the free case, these three expressions reduce to the same thing. 
 This can be easily seen as follows: suppose that the core is in virial equilibrium, fulfilling
\begin{equation} \label{virialcore}
2K_c + W_c=0.
\end{equation}
Next, we assume that the halo itself also fulfils his own virial equilibrium, i.e. 
\begin{equation} \label{virialhalo}
2K_h + W_h = 0.
\end{equation}
Of course, we might question in which sense it is meaningful to assume separate virial equilibrium, for the core and for the halo.  In practice, the above relationships will only hold approximately, especially for the halo which takes a longer time to virialize during which time the core might have already virialized.  
From (\ref{virialcore}), we have
\begin{equation}
\frac{K_c}{M_c} = -\frac{1}{2}\frac{W_c}{M_c},
\end{equation}
and (\ref{virialhalo}) implies
\begin{equation}
\frac{K_h}{M_h} = -\frac{1}{2}\frac{W_h}{M_h}.
\end{equation}
Combining the above relationships, equation (\ref{core_halo}) and using expressions for the potential energy,
\begin{equation}\label{ws}
W_c = -C^{(c)}_{grav}\frac{GM_c^2}{R_c},\ \ \ \ \  W_h = -C^{(h)}_{grav}\frac{GM_h^2}{R_h},
\end{equation}
with positive constants $C^{(c)}_{grav}$, $C^{(h)}_{grav}$ of order O(1) which depend upon the core and halo profiles, respectively,
this yields
\begin{equation}\label{fuzzy_ch}
C^{(c)}_{grav}\frac{M_c}{R_c} \simeq C^{(h)}_{grav} \frac{M_h}{R_h}.
\end{equation}
Observe that the above expression is equivalent to \eqref{vcvh}. 
Despite this "coincidence", we must emphasize that it cannot be expected to be true more generally, once contributions from extra terms (as is the case of a self-interaction between particles) are added to the system (compare \eqref{Energy_em} or \eqref{kinetic_em} with $v_c$ in \eqref{vcvh}, for example). 

On the other hand, in \citep{sc13} a different result was reported. In their simulations, non-cosmological but fully-virialized, they obtain a core-halo mass relation in the form 
\begin{equation}
    M_c\sim M_h^{5/9}.
\end{equation}
\citep{sc13} also reported an empirical relation for their results, given by
\begin{equation}\label{energy_mocz1_e}
    M_c\simeq 2.6\left(\frac{|E_h|}{(mG/\hbar)^2}\right)^{1/3},
\end{equation}
which can be also re-expressed, using the virial theorem \eqref{virial} in the free case as
\begin{equation}
    M_c\simeq 2.6\left(\frac{K_h}{(mG/\hbar)^2}\right)^{1/3} = 2.6\left(\frac{|W_h|}{2(mG/\hbar)^2}\right)^{1/3}.
\end{equation}
It is worth mentioning that these results have been reproduced by more authors, even in a cosmological context (see for example \citep{moczext1}), different from those reported by Schive et al. However, as \citep{wf2} realised, the finding of Mocz et al. in equation \eqref{energy_mocz1_e} can be understood, if the condition $E_c\simeq E_h$ applies (compare for example equation \eqref{energy_mocz1_e} and \eqref{energy_mocz1}), which from the virial theorem in the free case should be equivalent to $K_h\simeq K_c$ or $W_h\simeq W_c$.  This suggests that the halos generated in Mocz's et al. simulations were dominated by the central soliton. Therefore, the most general scenario may mandate a more general relation given by
\begin{equation}\label{mvcmvh}
    M_c v_c^2 \sim M_h v_h^2,
\end{equation}
or in other words, that the square of the circular velocity at the core radius would differ by a factor of $M_h/M_c$ from the one that is measured at the halo radius. Notice that this last consequence would not be limited to assuming that the halos were mostly dominated by the soliton.


{Now, in this section we are interested in extending the core-halo mass relation to SFDM models with self-interaction. A question that immediately arises before we proceed concerns the correct relation upon which we shall build our extension. 
Since it appears as if more simulation work will be required to settle this question, we will go ahead and work out an extension for each of the reported core-halo mass relation. This way, we can also clearly see which analytic premises are the basis of each relation, reported either in \citep{sc4,sc10} or \citep{sc13}.}



To begin with, it is necessary to be able to obtain quantities related to the entire halo. For our purposes and the scales of interest, it is sufficient to assume a halo with approximately constant density. In that case, all the energies defined in \eqref{Et} and \eqref{energies} can be expressed as:
\begin{subequations}
\begin{equation}
    E_t(R) = -\frac{3 G}{10}\frac{M^2(R)}{R}-\frac{3g}{8m^2\pi}\frac{M^2(R)}{R^3},
\end{equation}
\begin{equation}
    K_t(R) = \frac{3 G}{10}\frac{M^2(R)}{R}-\frac{9g}{8m^2\pi}\frac{M^2(R)}{R^3},
\end{equation}
\begin{equation}
    W_t(R) = -\frac{3 G}{5}\frac{M^2(R)}{R},
\end{equation}
\begin{equation}
    U_{SI,t}(R) = \frac{3g}{4m^2\pi}\frac{M^2(R)}{R^3},
\end{equation}
\end{subequations}
where in the above expression we have integrated from 0 to a given R. 
If we use the definition of the virial mass of the halo as $M_h = 4\pi R_h^3\rho_{200}/3$, where $R_h$ is the radius at which the mean density $\rho_{200}$ within such a radius is $200$ times larger than the background density, from the above expressions follows:
\begin{subequations}
\begin{equation}\label{E_en_const}
    E_h = -\left[\frac{ 3 }{10}\frac{GM_{crit}^{1/3}}{R_{crit}}M_h^{5/3}\pm\frac{1}{4}\frac{GM_{crit} }{R_{crit}}{M_h}\right],
\end{equation}
\begin{equation}\label{k_halo}
    K_h = \left[\frac{ 3 }{10}\frac{GM_{crit}^{1/3}}{R_{crit}}M_h^{5/3}\mp\frac{3}{4}\frac{GM_{crit} }{R_{crit}}{M_h}\right],
\end{equation}
\begin{equation}
    W_h = -\frac{ 3 }{5}\frac{GM_{crit}^{1/3}}{R_{crit}}M_h^{5/3},
\end{equation}
\begin{equation}
    U_{SI,h} = \pm\frac{1}{4}\frac{GM_{crit} }{R_{crit}}{M_h},
\end{equation}
\end{subequations}
where we have defined the quantity $M_{crit}\equiv 4\pi \rho_{200} R_{crit}^3/3$ and we have used again the critical radius $R_{crit} = \sqrt{6|g|/(4\pi G m^2)}$. From these last expressions we can already begin to infer several possible consequences regarding our possible extensions. {On the one hand, we can see that if the total halo is in the TF regime ($K_h\simeq 0$), from equation \eqref{k_halo} we arrive at $R_h\simeq R_c^{(TF)}$. Of course, this is unfavourable because it implies that we have one size for all halos in the Universe, and halos would effectively be limited to solitonic cores in the TF regime.}


{Therefore, any possible extension for models with self-interaction that pretends to maintain a NFW-like asymptotic exterior would necessarily have to consider a kinetic term, different from zero, to describe the complete halo. Since SFDM is expected to behave like CDM at large scales (i.e. scales much larger than either the de Broglie wavelength in the free case, or much larger than the TF radius in the TF regime), it should be true that at sufficiently large galactic scales, CDM should be recovered, suggesting that these NFW envelopes should be also found, even in the TF regime.} 
On the other hand, observe that the total energy of the halo can be expressed as
\begin{equation}
    E_h\sim -\left(\frac{3}{10}R_h^5\pm \frac{1}{4}R_{crit}^2 R_h^3\right), 
\end{equation}
so if we demand that $R_h\gg R_{crit}$, we can express the total energy of the system as
\begin{equation}
    E_h \simeq -\frac{ 3 }{10}\frac{GM_{crit}^{1/3}}{R_{crit}}M_h^{5/3}.
\end{equation}
Notice that the above expression results in $E_h\sim W_h \sim GM_h^2/R_h$, even if self-interaction is allowed for the SFDM particles. On the other hand, from \eqref{Energy_em}, the energy for the soliton would be always in the range
\begin{equation}
    \frac{1}{4\sqrt{2\pi}}\left(\frac{GM^2_c}{R_c}\right)\leq {|E_c|} \leq \frac{1}{3\sqrt{2\pi}}\left(\frac{GM^2_c}{R_c}\right),  \ g>0,\nonumber
\end{equation}
\begin{equation}
    \frac{1}{6\sqrt{2\pi}}\left(\frac{GM^2_c}{R_c}\right)\leq {|E_c|} \leq \frac{1}{4\sqrt{2\pi}}\left(\frac{GM^2_c}{R_c}\right),  \ g<0.\nonumber
\end{equation}
and similar relations for $W_c$, meaning that $|E_c|\sim |W_c|\sim GM_c^2/R_c$ in all cases. With this simple analysis we can conclude that the extensions that follow from the relations \eqref{vcvh} from the results of Schive et al., or \eqref{mvcmvh}\footnote{Observe that by adopting this relation we are also assuming that the results obtained by Mocz et al. are general and apply in cases for which the mass of the central soliton is much smaller than the mass of the total halo. } from the results of Mocz et al. should be sufficient to capture realistic results for our self-interaction models. 
\\$ $ 

\textit{Extending the core-halo mass relation, using the results of \citep{sc4,sc10}} \\

\begin{figure}[h!]
    \centering
    \includegraphics[width=3.4in]{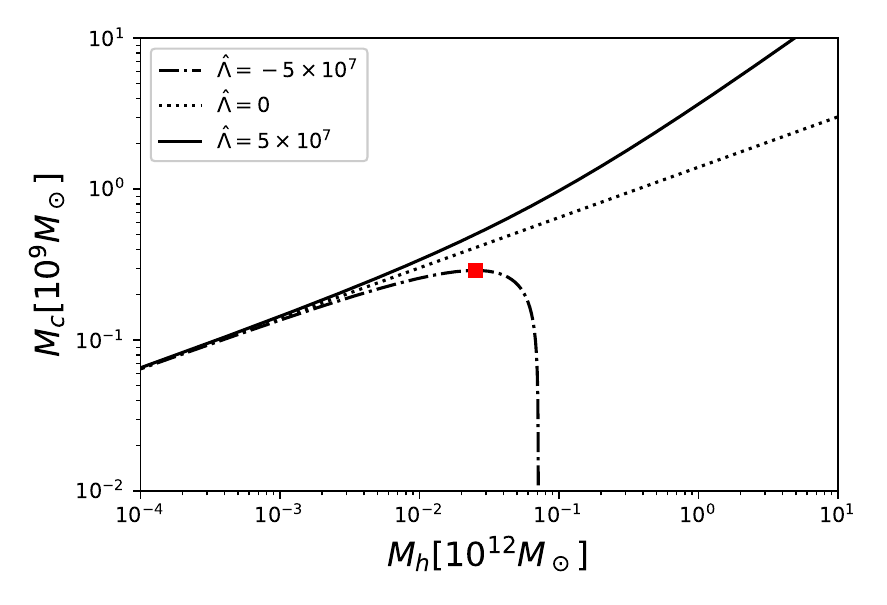}
        \includegraphics[width=3.4in]{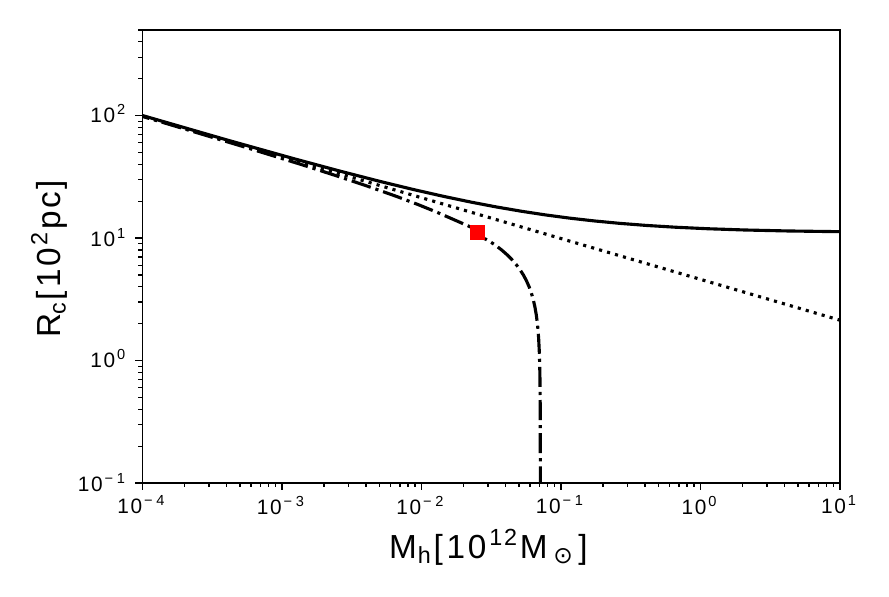}
\includegraphics[width=3.4in]{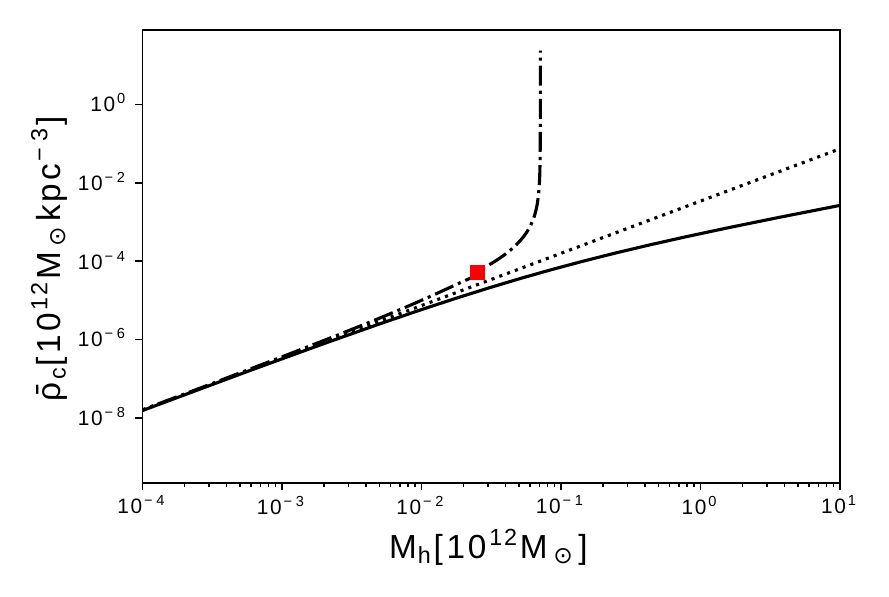}
        \caption{\footnotesize{Mass, radius and mean density for a soliton in self-interacting SFDM halos, in terms of total halo mass for our Schive et al. extension. For the following cases: repulsive self-interaction
($\hat\Lambda= 5 \times 10^7$;
solid curves), attractive self-interaction ($\hat\Lambda = - 5 \times 10^7$;
dot-dashed),
and no self-interaction ($\hat\Lambda = 0$; dotted).   The red-squares
labelling the attractive
case curves correspond to $M_c,max$.}}
    \label{fig:massvslam2}
\end{figure}

{Now, let us study the consequences arising once we assume equation \eqref{vcvh} as the correct extension for the core-halo mass relation in SFDM. While this work was in progress, an extension was also considered in \citep{chavanis2019core}, modelling the total halo with a generalized GPP system \citep{chavanis2018derivation}, obtaining that the total halo could be understood as a central soliton with an effective isothermal exterior. Nevertheless, we proceed to present here our own extension by following the procedures that we have applied so far. For this purpose, we assume that the core-halo mass relation, that we seek, is based upon the condition 
\begin{equation}\label{extension_schive}
    \frac{GM_c}{R_c}\simeq D_h\frac{GM_h}{R_h},
\end{equation}
where $D_h$ is a constant that must be fixed by numerical simulations. On the other hand, it is not difficult to rewrite the $M_c-R_c$ relation for the Gaussian ansatz as
\begin{equation}\label{M_c_core}
    M_c = (3\sqrt{2\pi})^{1/2}\frac{m_{pl}^2}{m}\frac{\sqrt{\frac{GM_c}{R_c}}}{c}\sqrt{1+\sqrt{\frac{2}{\pi}}\hat\Lambda\frac{\left(\frac{GM_c}{R_c}\right)}{c^2}}.
\end{equation}
We can express the above equation in terms of halo quantities using equation \eqref{extension_schive}. 
If additionally we replace $R_h = (3 M_h/4\pi\rho_{200})^{1/3}$, we obtain that the mass of the core can be expressed in terms of the mass of the complete halo as:
\begin{equation}
    M_c =\frac{m_{pl}^2}{m} \sqrt{3\sqrt{2\pi}\hat D_h M_h^{2/3}\left(1+\sqrt{\frac{2}{\pi}}\hat\Lambda\hat D_hM_h^{2/3}\right)}.
\end{equation}
where in the above expression we have defined $\hat D_h \equiv(4\pi\rho_{200}D_h^3 G^3/(3c^6))^{1/3}$.
Finding the numerical value of $\hat D_h$ by matching the above expression with $\hat\Lambda=0$ and the core-halo mass relation in the free case by Schive et al. \eqref{M_c}, we finally arrive at the core-halo mass relation in fiducial units:
\begin{equation}\label{mcmhchavanis1}
    M_{c,7} = \frac{1.4\times 10^2 M_{h,12}^{1/3}}{m_{22}}\sqrt{1+\hat\Lambda ({1.16\times 10^{-7}M_{h,12}^{2/3}})}.
\end{equation}

To understand the consequences that follow from the above result, we plot in figure \ref{fig:massvslam2} $M_c$ (top panel), $R_c$ (middle panel) and $\bar\rho_c\equiv 3M_c/(4\pi R_c^3)$ (bottom panel) as a function of $M_h$ setting $m_{22} = 1$. Observe that from this extension we obtain that in the attractive scenario, there is a critical halo mass:
\begin{equation}\label{m_hcrit_schive}
M_{h,12}^{(crit)} = \left(\frac{4.29\times 10^{6}}{|\hat\Lambda|}\right)^{3/2},
\end{equation}
at which the central soliton arrives at its maximum possible mass
\begin{equation}\label{critical_fiducial}
    M_{c,7}^{max} = \frac{2.05\times 10^{5}}{m_{22}|\hat\Lambda|^{1/2}},
\end{equation}
(red-square in the plot).  It is not difficult to show that the above expression coincides with \eqref{maximum_mass_neg} once rewritten in appropriate units. On the other hand, when $\hat\Lambda>0$ and $\hat\Lambda (1.16\times 10^{-7}M_{h,12}^{2/3})\gg 1$, we arrive at the TF regime for the central soliton profile, resulting in a core-halo mass relation in the form:
\begin{equation}\label{TF_e}
    M_{c,7}\simeq 4.78\times 10^{-2}\frac{\sqrt{\hat\Lambda}}{m_{22}}M_{h,12}^{2/3}.
\end{equation}
This core-halo mass relation in the TF regime is in agreement with the one obtained in \citep{chavanis2019core}.}
\\ 

\textit{Extending the core-halo mass relation, using the results of \citep{sc13}} \\

\begin{figure}[h!]
    \centering
    \includegraphics[width=3.4in]{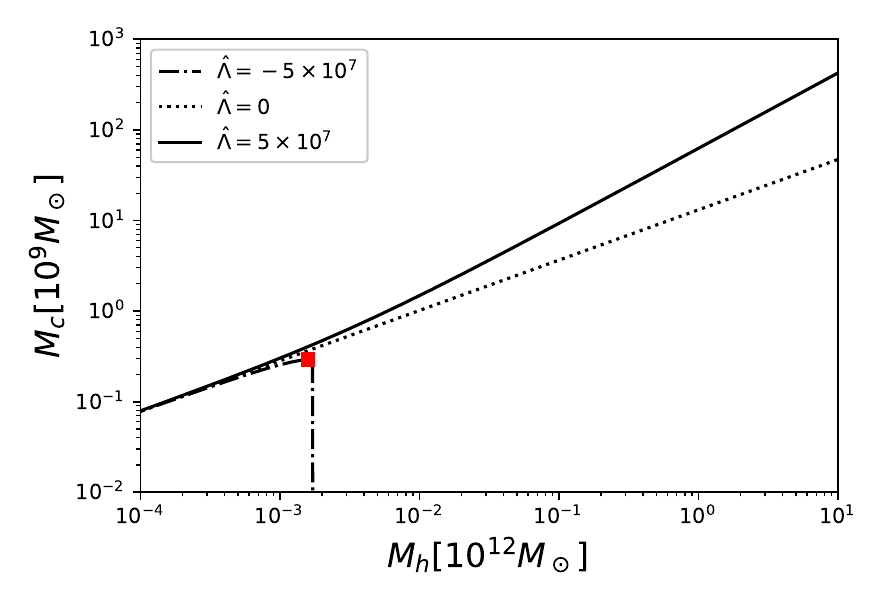}
       \includegraphics[width=3.4in]{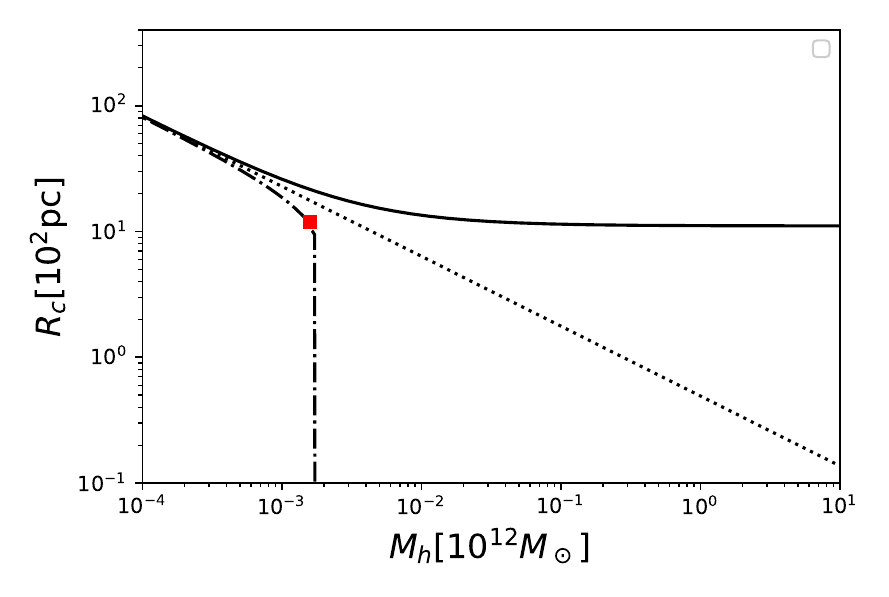}
\includegraphics[width=3.4in]{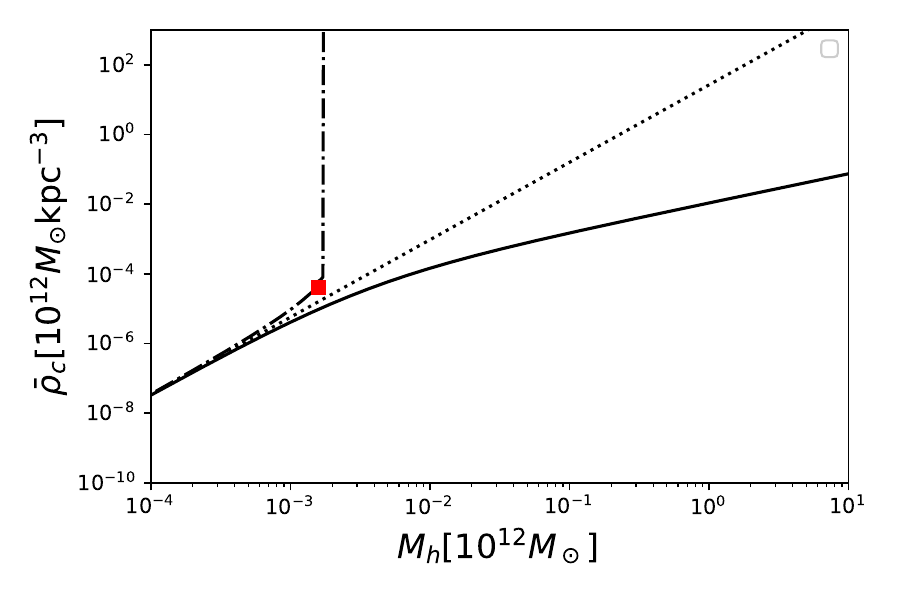}
        \caption{\footnotesize{Same as Fig. \ref{fig:massvslam2}, except for our Mocz etal.
extension, instead. }}
    \label{fig:massvslam3}
\end{figure}

For this extension we shall consider that the condition that should correctly describe core-halo quantities is
\begin{equation}
    \frac{GM^2_c}{R_c}\simeq C_h\frac{GM^2_h}{R_h},
\end{equation}
where, as before, $C_h$ is a constant that must be fixed by numerical results. Multiplying equation \eqref{M_c_core} by $M_c$ on both sides, using the above expression, and $R_h = (3 M_h/4\pi\rho_{200})^{1/3}$, it is easy to see that the core-halo mass relation in this approximation is given by the expression
\begin{equation}\label{Mc2_mocz}
    M_c^2 =\frac{m_{pl}^2}{m} \sqrt{3\sqrt{2\pi}\hat C_h M_h^{5/3}\left(M_c+\sqrt{\frac{2}{\pi}}\hat\Lambda\hat C_hM_h^{5/3}\right)},
\end{equation}
where $\hat C_h \equiv(4\pi\rho_{200}C_h^3 G^3/(3c^6))^{1/3}$. Similarly than in our extension above, we obtain the numerical value of $\hat C_h$ by matching the above expression with numerical simulation results. In order to do so we will proceed using equation \eqref{energy_mocz1_e} and noticing that once we use \eqref{E_en_const} in the free case, it can be rewritten as:
\begin{equation}
    M_c\simeq 2.6\left[\frac{m_{pl}^4}{m^2c^2}\frac{3}{10}\left(\frac{4\pi\rho_{200}}{3}^{1/3}\right)^{1/3}M_h^{5/3}\right]^{1/3}.
\end{equation}
If for consistency we use the current mean density of the Universe $\rho_{b} = 1.5\times 10^{-7} M_\odot\rm{pc^{-3}}$, such as $\rho_{200}=200\rho_b$, we obtain that the core-halo mass relation obtained in Mocz et al. simulations should be roughly given by
\begin{equation}\label{M_c_mocz}
    M_{c,7}\simeq 1.31\times 10^3 \frac{M_{h,12}^{5/9}}{m_{22}^{2/3}}.
\end{equation}
Comparing this
with \eqref{Mc2_mocz} in the free case, we obtain finally that the core-halo mass relation with self-interaction corresponds in fiducial units to
\begin{equation}\label{mc_mh_mocz}
    M_{c,7}^2 \simeq \frac{4.75\times 10^4 M_{h,12}^{5/6}}{m_{22}}\sqrt{M_{c,7}+\hat\Lambda(1.34\times 10^{-2} M_{h,12}^{5/3})}.
\end{equation}
Again, we plot $M_c$ (top panel), $R_c$ (middle panel) and $\bar\rho_c $ (bottom panel) as a function of $M_{h,12}$ for this case in figure \ref{fig:massvslam3}. Observe that in all cases, this extension results in larger masses for the central soliton than in the Schive et al. extension. This conclusion was also pointed out by Mocz et al. for their results of the free case.  
In addition, similarly to the other extension, there is a critical total halo mass for the attractive scenario at which the central soliton arrives at its maximum possible mass:
\begin{equation}\label{m_hcrit_mocz}
    M_{h,12}^{(crit)} = \frac{1.34\times 10^4}{(m_{22}|\hat\Lambda|^{3/2})^{3/5}},
\end{equation}
(red-square in the plot). Of course, if we replace this mass in equation \eqref{mc_mh_mocz}, we arrive at the same result than in equation \eqref{critical_fiducial}, which, as we explained before,  is just the same as \eqref{maximum_mass_neg} but rewritten in our fiducial units.
On the other hand, when $\hat\Lambda>0$ and $\hat\Lambda(1.34\times 10^{-2}M_{h,12}^{5/3})\gg M_{c,7}$, we arrive at the TF regime, in which the core-halo mass relation in this limit is given by
\begin{equation}\label{TF_mocz}
    M_{c,7}\simeq 74.15\left(\frac{\sqrt{\hat\Lambda}}{m_{22}}\right)^{1/2}M_{h,12}^{5/6}.
\end{equation}

\section{\label{sec:basiceq}Astrophysical consequences for SFDM with self-interaction}
    
A fundamental question that arises in the SFDM model concerns the values of its free parameters. Many constraints have been derived already, using cosmological and astrophysical data. In this section, we first review some of the most representative results obtained in the literature and apply them to our extensions, i.e. we shall confront our extensions obtained, using the simulation results of Schive et al. and those of  Mocz et al., respectively. The main way we decided to explore the parameter space of SFDM and our core-halo mass extensions is by exploring the possible SMBH formation in each scenario, although we will also comment on other consequences that will arise once we confront the parameter region that has been studied by other observational data. 

As we already mentioned, SMBHs with masses in the range $M_{SMBH}\simeq 10^6-10^{10}~M_\odot$ have been found in almost all large massive galaxies, while this is not the case for the smallest ones, like small dSphs. { However, BHs of masses around $10^6 M_\odot$ have been found in some dwarf galaxies, e.g. \citep{ahn2017detection}, while \citep{dsphsmbh} have reported BHs in dwarf galaxies with smaller masses, $10^4-10^5 M_\odot$, albeit this non-dynamical mass estimate is very much uncertain\footnote{Also, recently a black hole has been detected in the intermediate mass range with a mass of $\sim 150 M_\odot$, which would be generated by the merger product of 2 smaller black holes \citep{abbott2020gw190521}. Of course, our intention in this work is only to explain the SMBHs within galactic nuclei, so the formation of these objects is not covered by our scenario.}.} Thus, we could try to find scenarios in which the central solitons reach their maximum possible mass only for the most massive galactic halos (hosting the most massive galaxies).  {However, it is unclear at this point whether the core-halo mass relations found in current simulations of free SFDM - on which we based our extensions - remain valid for the most massive galactic-size halos, because those simulations were limited to small volumes. Therefore, it seems appropriate to apply the core-halo mass relations to the question of SMBH formation in halos whose critical mass for soliton collapse, $M_{h,12} = M_{h,12}^{(max)}$, does not exceed a certain limit, which we choose to set at 1 (in these units). We may think of these SMBHs as the ``seeds" for possibly even more massive black holes in the centers of the most massive galaxies.}

For consistency with observations, however, {the mass of these SMBH seeds would be expected to not exceed the mass of the least massive SMBHs found in galaxies, as e.g. in our own Milky Way. We could also assume that these SMBH seeds could be orders of magnitude smaller, { but that mass range would fall into the so-called intermediate mass range, which is still under debate, as mentioned above.}  Therefore, we will only focus on SMBH seeds that are still in the supermassive range.
More precisely, we will consider a fiducial mass range of such ``seed" SMBHs, equivalently to consider a range for the maximum mass of solitonic cores of $M_{c,7}^{max}\simeq 0.1-10$, whose collapse is supposed to form these SMBHs. }
Similarly, we need to take fiducial values for $M_{h,12}^{(crit)}$ such that these SMBHs are not formed for the least massive galaxies, such that they keep having  a stable soliton core in their centers, nor for very large galaxy/halo masses, as explained above. 
In analogy to the above description, we could expect that this critical mass of collapse corresponds to the minimum mass of a galactic halo in whose center the presence of a SMBH is expected, however, we decided again to be flexible, and we adopted as a conservative criterion in which galactic halos with masses in the range $M_{h,12}^{(crit)}\simeq 10^{-2}-1$ are the ones that will ``start" to possess a SMBH at their centre. In what follows, we will call the combination of these two ranges of fiducial parameters (when both are met simultaneously) as our ``ideal model", {where, of course, it is understood that we refer to this ideal model only in the context of the formation of these seeds in the supermassive range.}

\subsection{The free case ($\hat\Lambda = 0$)}

This is the best-studied case in the literature and for this scenario there are several constraints that have been found by different groups. We shall review only some representative constraints that have been obtained for this scenario. 

Using the hydrodynamical representation of the SFDM model, in \citep{free_const1} it was suggested that the quantum pressure of SFDM explains the offset between DM and ordinary matter in Abel 3827. For this purpose, they
required a mass $m_{22} \simeq 2\times 10^{-2}$. When the model is tested with the dynamics of dSphs -- Fornax and Sculpture --, in reference \citep{free_const2}, a mass
constraint of $m_{22} < 0.4$ at 97.5\% was obtained. The constraints which follow when the survival of the cold
clump in Ursa {Minor} and the distribution of globular clusters in Fornax is demanded, requires a
mass $m_{22} \sim 0.3 - 1$ \citep{free_const3}. Explaining the half-light mass in the ultra-faint
dwarfs fits the mass to be $m_{22} \sim 3.7 - 5.6$ \citep{free_const4}. The model
has also been constrained by observations of the reionization process. In \citep{free_const5}, using N-body simulations and demanding an ionized fraction of HI of 50\% by z = 8,
$m_{22} > 0.26$ was obtained. Finally, the Lyman-$\alpha$ forest flux power spectrum
demands that the mass parameter fulfils $m_{22} \geq 20 - 30$ \citep{free_const6,free_const7}.

Now, we shall try to explain the formation of SMBHs for free SFDM, in light of the discussion of the beginning of this section. For this case, it is sufficient to apply the results obtained by Schive et al. \eqref{M_c} and Mocz et al. \eqref{M_c_mocz}, respectively.

Observe that by equating the core-halo mass relation \eqref{M_c} and the critical mass of collapse \eqref{mmasfree} for a soliton in the free case, which in fiducial units reads:
\begin{equation}\label{crit_fiducial}
    M_{c,7}^{max} = \frac{8.46\times 10^4}{m_{22}},
\end{equation}
we should have that the maximum possible soliton mass is reached for a critical halo mass of
\begin{equation}\label{mc_crit_free}
    M_{h,12}^{(max)} \simeq 2.204\times 10^{8}.
\end{equation}
{This value exceeds by many orders of magnitudes even the halos around the most massive galaxies with $M_{h,12}\sim 10^2$.} Therefore, for SFDM without self-interaction, solitons will never collapse to form SMBHs. 
Interestingly, this result is independent of the mass of the SFDM particle.  {Therefore, we find that the core-halo mass relation due to \citep{sc4} implies that the formation of SMBHs by soliton collapse is not possible.}

\begin{figure}
    \centering
    \includegraphics[width=3.4in]{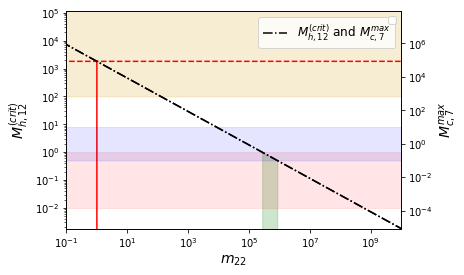}
    \caption{\footnotesize{$M_{h,12}^{(crit)}$ (left ``$y$"-axis) and $M_{c,7}^{max}$ (right ``$y$"-axis) as a function of $m_{22}$. {The blue and red regions correspond to our fiducial values $M_{c,7}^{max}=0.1-10$ and $M_{h,12}^{(crit)}=10^{-2}-1$, respectively, whereas the green region represents the parameter range in $m_{22}$ that fulfils the ``ideal model", i.e. as a result of the above fiducial choice.} We indicate in golden (upper part in plot) the range $M_{h,12}^{(crit)}>10^2$ of galactic halos whose mass range is excluded by observations.   (see the main text for more information).}}
    \label{fig:mass_free_crit}
\end{figure}

On the other hand, we can proceed in the same way with equation \eqref{M_c_mocz} and \eqref{crit_fiducial}. In this case, we obtain that the critical value at which the central soliton arrives at its maximum possible mass would be given by
\begin{equation}\label{mh_12_mocz}
    M_{h,12}^{(max)}\simeq \frac{1.81\times 10^3}{m_{22}^{3/5}},
\end{equation}
i.e. a different result than in the above procedure. {In fact, if we adopt the core-halo mass relation of \citep{sc13} as the correct one, we could have scenarios in which the central soliton in galaxies can collapse and form a SMBH for some values of the mass parameter of SFDM. In order to highlight what the values for the mass parameter must be in order to achieve soliton collapse}, in figure \ref{fig:mass_free_crit} we plotted $M_{h,12}^{(crit)}$ (equation \eqref{mh_12_mocz}; left ``$y$"-axis) and $M_{c,7}^{max}$ (equation \eqref{crit_fiducial}; right ``$y$"-axis) with a dot-dashed black-line as a function of $m_{22}$. To understand better how this figure should be read, let us focus on a special case, for example the case for which $m_{22} = 1$. In figure \ref{fig:mass_free_crit} we plotted this case with a red vertical line. Observe that in some point this line intersects with the dot-dashed black line. It is precisely at this intersection where we can talk about the critical masses $M_{h,12}^{(crit)}$ and $M_{c,7}^{max}$ that corresponds to this particular example. To know exactly what these masses are, the reader may just look at the horizontal dashed red line and the point at which it intersects both ``$y$"-axes, i.e. for the special case of $m_{22}=1$ we obtain that galaxies with a critical mass $M_{h,12}^{(max)}=1.81\times 10^3$ should posses a soliton with a mass that equals its maximum possible mass $M_{c,7}^{max}=8.46\times 10^4$. 

It is clear that this example is far away from belonging to our "ideal model". For this reason, we draw in blue and red the fiducial mass ranges of our "ideal model", i.e.  $M_{c,7}^{max} = 0.1-10$ and $M_{h,12}^{(crit)}=10^{-2}-1$, respectively. The first thing we can observe from these two regions is that they only intersect for a small region, {which corresponds to having collapse with critical masses of $M_{c,7}^{max} = 0.31-0.1$ once the mass of the galactic halo exceeds the corresponding critical values $M_{h,12}^{(crit)} = 1-0.5$.} This intersection coincide with values of the mass parameter in the range $m_{22} = 2.69\times 10^5 - 8.46\times 10^5$ (green region in the plot). This implies that we could meet our ``ideal model" only for that range of parameters, implying that if we were interested in explaining the possible formation of SMBHs, the range $m_{22} = 2.69\times 10^5 - 8.46\times 10^5$ would be favoured. Let us explain in more detail why we consider these masses as the ones that are favoured. Suppose our ``ideal model" is only partially fulfilled. For example, if we try to meet the condition of obtaining collapse, once we reach critical masses of $M_{h,12}^{(crit)}= 10^{-2}-1$, it would result in the formation of SMBH seeds with masses of $M_{c,7}^{max}\simeq 1.46\times 10^{-4}-3.15\times 10^{-1}$, so the mass of most of these SMBH seeds would be well below the mass of typical SMBHs found in galactic nuclei, and instead they would correspond to intermediate mass black holes. Since we do not consider the intermediate mass range in our scenario, the only masses that could account for such SMBH seeds would be the most massive ones, those that fall within the overlap region in our ``ideal model". Similarly, if instead we require to meet the condition of obtaining collapse for soliton masses $M_{c,7}^{max} = 0.1-10$, the mass of the galactic halos for which such collapse could be achieved is $M_{h,12}^{(crit)} = 0.5-8.07$, leading to a correspondingly rather large range of galaxy masses. Although these masses do exist in the Universe, we discard the most massive of them (the ones that are not covered by our ``ideal model"), because SMBHs have been seen even in less massive galaxies. So, if we expect to explain them through this mechanism, it would be necessary for them to be formed from smallest critical halo masses. 

Finally, we should compare our estimates for the particle mass parameter with the constraints that have been found by other groups. Once we make the comparison we can observe that our range of preferred values are completely in disagreement with most of the previous studies. {At best, the only agreement can be found with those that follow from reionization and the Lyman-$\alpha$ forest flux power spectrum. Despite this discrepancy, it should be taken into account that adopting values for the particle mass of SFDM that are as large as ours would imply that our model increasingly resembles standard CDM, so that the range of parameters that we are taking for our ``ideal model" can not be ruled out as of yet, if we accept the notion that CDM has not yet been ruled out. In that case, of course, SFDM with such model parameters would not resolve the CDM small-scale crisis, either, but instead, it could help to explain SMBH formation (which CDM does not), which makes it attractive as an alternative for the dark matter in the Universe.}
{By the same token, our model can help us to turn the question around and demand to put some extra bounds on the particle mass parameter of SFDM. It is clear that we should avoid to make SMBHs of the wrong
mass and/or for the wrong halo masses. As we have mentioned before, if SMBHs with masses smaller than supermassive were generated (in the intermediate range), there would be a problem because these BH populations have not yet been firmly detected. That way, we could partially discard this parameter region ($m_{22}>8.46\times 10^5$). On the other hand, if we have soliton collapse for galactic halos that are more massive than our fiducial model, we could also rule out that parameter region since, as we mentioned earlier, this is not desirable if we want to use this mechanism to explain the presence of SMBHs in galaxies with smaller masses. If we assume that typically the most massive galactic halos have a mass of the order of $M_{h,12}\simeq 10^2$, the above condition ruled out the region $m_{22} = 1.25\times 10^2-2.69\times 10^5$. Finally, the \textit{non-formation} of SMBHs by this mechanism would also give us a region of allowed parameters for this model.} If we demand that for a mass $M_{h,12}^{(crit)}=10^2$ the critical mass for collapse of the soliton has not yet been reached (we mark this region in golden in figure \ref{fig:mass_free_crit}), we would have that the mass parameter of the model should meet the condition $m_{22} \leq 1.25\times 10^2$. {As such, this last constraint does agree with all those that have been previously reported.}

\subsection{Repulsive case ($\hat\Lambda>0$)}

This is a scenario that has been also extensively studied in the literature and its free parameters have been fitted using different observations. Usually, the strong self-interaction regime is considered, because of simplicity, and in this case, it is the ratio $g/(m^2c^4)$ which is subject to constraints. Observe that from \eqref{parameters} we have
\begin{equation}\label{self_}
    \hat\Lambda = 1.54m_{22}^2\times 10^{37}\left(\frac{g}{m^2c^4}\right)\frac{eV}{cm^3},
\end{equation}
i.e. we can likewise constrain $\hat\Lambda$. In this section we shall use only the constraints that have been found in the strong self-interaction regime, for one thing, because the bounds are stronger and may hold for weak self-interaction, as well. It is these bounds which will be put into context to our results.

The first constraint, applicable to all candidates for dark matter, refers to the fact that by the redshift of radiation-matter equality $z_{eq}$, they must all be non-relativistic, i.e. behaving like a pressureless fluid. It is well known that a scalar field with an arbitrary potential $V(\varphi)$\footnote{Here $\varphi$ is the scalar-field that appears in the Klein-Gordon equation and is related to $\psi$ via equation \eqref{twoscalars}.} will have a varied dynamics during its cosmological evolution. In particular, the dynamics of a self-interacting SFDM candidate with a repulsive self-interaction has been studied previously and can be briefly summarized as follows \citep{sf5,self-const1,cosmo_self}: After inflation, the SFDM energy density behaves either like a cosmological constant ($\rho_\varphi\propto a^0$), or a stiff fluid ($\rho_\varphi\propto a^{-6}$), depending upon whether SFDM is effectively a real or complex field, respectively. This behavior of SFDM is rooted in the \textit{slowly oscillating} phase and is characterized by $\omega^2\equiv 2c^2dV/d|\varphi|^2\ll H^2$. However, in its \textit{fast oscillating} regime ($\omega^2\gg H^2$), there are two possible branches for SFDM \citep{self-const1, self-const2}: for \textit{weak self-interaction}, SFDM transitions from the stiff phase to the pressureless phase without having a radiation-like behavior in between. This happens, because the first term in the scalar field potential \eqref{potential} dominates over the second term at the moment of transition from slow to fast oscillation. On the other hand, for \textit{strong self-interaction}, SFDM transitions from the stiff phase to a radiation-like phase, before behaving like a pressureless fluid. Demanding that at $z_{eq}$, SFDM should be in its pressureless phase implies a constraint as follows \citep{self-const1}:
\begin{equation}\label{self_const3}
    \frac{\hat\Lambda}{m_{22}^2}\leq 6.18\times 10^{20}.
\end{equation}
This result represents an upper bound for the self-interaction parameter, including the weak self-interacting regime. This last result is also independent of whether SFDM is real or complex, given that the strong and the weak regimes are applicable to both cases. Hence, the above result is applicable to all self-interacting SFDM models with a repulsive self-interaction.  
 
On the other hand, the repulsive SFDM model has been also probed by studying the effective number of relativistic degrees of freedom during Big Bang nucleosynthesis (BBN), $N_{eff,BBN}$ \citep{self-const1}. The analysis was performed in the strong self-interacting regime for a complex SFDM candidate, and it was shown that this scenario can be made in accordance with BBN bounds. Using the allowed $1\sigma$-band on $N_{eff,BBN}$ at that time, it was shown that the ratio $g/(m^2c^4)$ must fulfill an upper {and} a lower bound. However, if the lower bound of the $1\sigma$-band on $N_{eff,BBN}$ is relaxed, i.e. if BBN is considered in accordance with the standard value of $N_{eff} = 3.046$, then the ratio $g/(m^2c^4)$ can be much smaller than the above upper bound suggests, as long as the boson mass $m$ fulfils a corresponding lower bound constraint, which ensures that the stiff-like era ends at an early enough time. This analysis has been extended in \citep{self-const3}, to include a scenario where the stochastic gravitational wave background (SGWB) from inflation could be amplified, as a result of the stiff-like behavior of SFDM in the very early Universe, after reheating, when SFDM dominates the mean energy density in the Universe. In this case both, SFDM and the inflationary SGWB, contribute to $N_{eff,BBN}$. The modified bounds which result effectively shrink the available parameter space of complex SFDM further, but in doing so the SGWB is boosted to a level where it can be potentially observed by LIGO (see \citep{self-const3}). However, if the stiff phase ends early enough, such that the SGWB remains negligible, the lower and upper bounds on $g/(m^2c^4)$ are determined basically again by demanding that SFDM fulfills BBN bounds. An updated value for $N_{eff,BBN}$
has been used in \citep{self-const3} to derive newer bounds for this case, as well. Using \eqref{self_}, the corresponding bounds read as
\begin{equation}\label{self_const4}
    3.55\times 10^{19}\leq \frac{\hat\Lambda}{m_{22}^2} \leq 6.33\times 10^{20}.
\end{equation}
Interestingly, if we use the above parameters in the TF radius \eqref{TFradius}, it turns out to be of order $R^{(TF)}\sim \rm{kpc}$.

Similarly to the previous subsection, we will try to see if it is possible to obtain scenarios for SMBH formation in this case. For this, it will be necessary to use each of our extensions of the core-halo mass relations that we derived in the previous section, that is, we shall probe our extensions \eqref{mcmhchavanis1} and \eqref{mc_mh_mocz}, respectively. We will consider that the central soliton reaches its maximum possible mass by assuming the already known numerical result \eqref{mass_c_pos}, which in fiducial units reads \begin{equation}\label{eq_97}
    M_{c,7}^{max} = 2.94\times 10^4\frac{\sqrt{\Lambda}}{m_{22}}
\end{equation}
On the other hand, it is not difficult to anticipate that this result is valid for those solitons that are well within the TF regime (as we analyzed when we compared our Gaussian ansatz with the numerical results in section \ref{relat_corr}). For this purpose, we will apply our extensions in the TF limit, i.e. using equations \eqref{TF_e} and \eqref{TF_mocz}.

Let us start by analyzing the extension that is based upon the results from Schive et al. If we compare \eqref{TF_e} and \eqref{eq_97}, we find that the central soliton will collapse above a critical halo mass:
\begin{equation}
    M_{h,12}^{(crit)}\simeq 4.82 \times 10^8.
\end{equation}
This is close to the one that we obtained in the free SFDM case, \eqref{mc_crit_free}, and just like there, this quantity does not depend upon the free parameters of the SFDM particle, {$m$ and $\lambda$}. 
Again, through this extension it is not possible to form SMBHs via soliton collapse, for the critical halo mass is many orders of magnitudes too high.  

\begin{figure}
    \centering
    \includegraphics[width=3.4in]{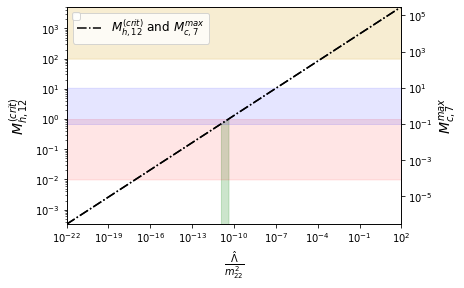}
    \caption{\footnotesize{$M_{h,12}^{(crit)}$ and $M_{c,7}^{max}$ as a function of ${\hat\Lambda}/m_{22}^2$. The blue, red, golden and green regions indicate the same meaning as in in figure \ref{fig:mass_free_crit}.}}
    \label{fig:mass_selfpos_crit}
\end{figure}

Now, let us study the extension that is based upon the results which follow from Mocz et al. If we compare \eqref{TF_mocz} and \eqref{eq_97} we obtain
\begin{equation}\label{eq_99}
M_{h,12}^{(crit)}\simeq 1.31 \times 10^3 \left(\frac{ \sqrt{\hat\Lambda}}{m_{22}}\right)^{3/5}.
\end{equation}
In figure \ref{fig:mass_selfpos_crit} we plotted $M_{h,12}^{(crit)}$ (equation \eqref{eq_99}; left ``$y$"-axis) and $M_{c,7}^{max}$ (equation \eqref{eq_97}; right ``$y$"-axis) as a function of $\hat\Lambda/m_{ 22}^2$. Similarly than in the free case, we draw {in blue and red our choice of fiducial ranges of $M_{c,7}^{max}= 0.1-10$ and $M_{h,12}^{(crit)} = 10^{-2}-1$, respectively.}
As before, we end up with a small range of SFDM parameters that can fulfil our ``ideal model" (we marked them in green), which are $\hat\Lambda/m_{22}^2 = 1.16\times 10^{-11}-4.06\times 10^{-11}$. Only this region of parameters would be favoured for possible SMBH formation.

Finally, we need to confront our estimate for SMBH formation and the previous constraints in the literature. We can see that our estimate is in accordance with the constraints in \eqref{self_const3} but would be in disagreement with BBN constraints, {unless the latter are relaxed by considering the limit of very weak self-interaction}. However, somewhat similar to the free case, this scenario can not be ruled out per se, since this region of parameters corresponds to having a small TF radius\footnote{As we have seen in \eqref{gaussian}, the TF radius can be expressed, using \eqref{parameters} in \eqref{TFradius}, as $R^{(TF)}\propto \frac{\hat\Lambda}{m_{22}^2}$, i.e. a small value of the ratio $\hat\Lambda/m_{22}^2$ implies a small TF radius.}, meaning the model resembles standard CDM. On the other hand, following the same description we did in the free case, we can rule out the region of parameters that pertains to $\hat\Lambda/m_{22}^2<1.16\times 10^{-11}$ to avoid the formation of small BH seeds, and $\hat\Lambda/m_{22}^2 = 4.06\times 10^{-11}-1.89\times 10^{-4}$ to avoid the formation of SMBHs for large halo masses. The requirement of \textit{non-formation} of SMBHs imposes the condition $\hat\Lambda/m_{22}^2\geq 1.89\times 10^{-4}$. This last constraint is in agreement with all the previous constraints that we reviewed. If we were to adopt this constraint, SMBHs would not form in this scenario, and the core-halo mass relation for this model would be given by equations \eqref{TF_e} or \eqref{TF_mocz}.

Remark: Observe that we require very small values for $\hat\Lambda/m_{22}^2$ in order to explain the possible formation of SMBHs in this model, so we might think that these values should not necessarily be within the TF regime. However, from equation \eqref{R_cs} we know that the TF regime is reached for $6\hat\Lambda (M_{c,7}m_{22}/(5.038\times 10^{5}))^2\gg 1$. If we replace $M_{c,7}$ by $M_{c,7}^{max}$, this expression imposes the condition:
\begin{equation}
    \frac{\hat\Lambda}{m_{22}^2}\gg \frac{17.14}{m_{22}^2}.
\end{equation}
{As long as this condition is fulfilled, the model will be in the TF regime.}


\subsection{Attractive case ($\hat\Lambda <0$)} 

This scenario is the least-studied in the literature in the context of halo formation and dynamics, given that the  self-interaction term is so small that its contribution is often ignored. Therefore, this scenario is usually modelled effectively in the free-field regime, so the constraints of the mass parameter that are obtained in the free-field model are usually shared for this scenario as well. Despite that, there have been few works that have tried to incorporate the effects of an attractive self-interaction, so there are few observations that are used to put limits on this parameter. The constraints that are found for the self-interacting parameter are usually imposed on the parameter $\lambda$, however, we can re-express them in terms of $\hat \Lambda$ by using \eqref{parameters} as
\begin{equation}\label{hatlamlam}
    \hat \Lambda = 5.93\times 10^{98} \frac{\lambda}{m_{22}^2}.
\end{equation}

In \citep{chavanis} it was studied the evolution of the background Universe of this model. If the SFDM candidate is an ultra-light axion-like particle  ($m_{22} \sim 1$) -- a pseudo Nambu-Goldstone boson generated by a spontaneously broken global $U(1)$ symmetry --, it was suggested in \citep{axion_like} that these particles {should} be generated during the inflationary epoch in order to avoid observational constraints from Planck data, due to topological defects. In the same work, it was argued that by demanding that the total DM observed today is composed of these ultra-light axions, they should have a self-interaction parameter\footnote{The self-interaction parameter is obtained as $\lambda = m^2/f^2$, where $f$ is the axion-decay constant. For an ultra-light axion, the decay constant is of order $f\sim 10^{16}GeV$.} 
$|\hat\Lambda| \sim 5.93\times 10^{4}$, although, they also showed that the value of this self-interaction term can increase as soon as the value of the mass parameter also increases. These axion-like particles have also been studied in the context of type IIB orientifold compactifications in string theories, resulting in the possibility of obtaining stronger self-interactions 
$|\hat\Lambda| \sim 5.93\times 10^{12}$ for the case $m_{22}\sim 1$ \citep{lowf}\footnote{Notice that in terms of the parameter $\lambda$, these values for the self-interaction are extremely small, which corroborate the fact that self-interaction is usually ignored.}. Astrophysical considerations can lead to further novel constraints, e.g. the soliton with the maximum mass and smallest radius was matched to the smallest galaxy then known -- Willman I -- in \citep{supermassive}. By demanding that the halo of Willman I is dominated by the self-interacting soliton close to its maximum possible mass, the SFDM parameters were constrained to be $m_{22} = 193$ and $|\hat\Lambda| = 3.25\times 10^{8}$. 
In that case, the critical mass for collapse of a soliton should be close to the Willman I mass, i.e. $M_{c,7}^{max}\sim 0.1$, which is in agreement with one of our requirements of our ``ideal model" (namely the one that requires collapse once the mass of the central soliton exceeds $M_{c,7}^{max}=0.1-10$). However, this estimate does not fulfil the other condition of obtaining collapse once $M_{h,12}^{(crit)}=10^{-2}-1$.

\begin{figure}[h!]
    \centering
    \includegraphics[width=3.4in]{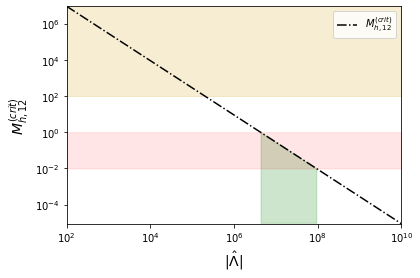}
    \includegraphics[width=3.4in]{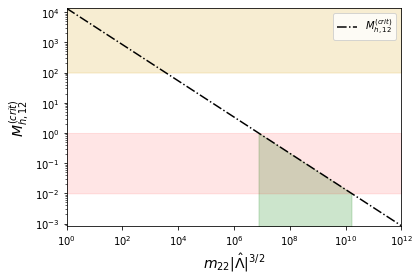}
4\includegraphics[width=3.5in]{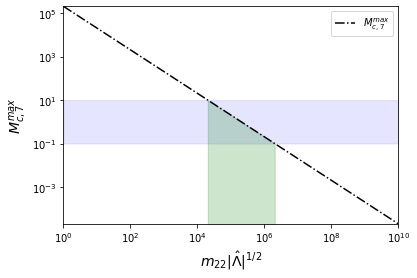}
    \caption{\footnotesize{(Top) $M_{h,12}^{(crit)}$ as a function of $|\hat\Lambda|$ for our Schive et al. extension. (Middle) $M_{h,12}^{(crit)}$ as a function of $m_{22}|\hat\Lambda|^{3/2}$ for our Mocz et al. extension.
    (Bottom) $M_{c,7}^{max}$ as a function of $m_{22}|\hat\Lambda|^{1/2}$. The blue, red, golden and green regions indicate the same meaning as in figure \ref{fig:mass_free_crit}.}}
    \label{fig:mass_selfneg_crit}
\end{figure}

Now, in this subsection we shall proceed in the same way as in the previous scenarios, i.e. we shall try to find the region of parameters where our ``ideal model" can be fulfilled. The first thing we can see is that, different from the previous cases, once we compare the critical mass $M_{h,12}^{(crit)}$ \eqref{m_hcrit_schive} or \eqref{m_hcrit_mocz} and the maximum possible mass for the soliton $M_{c,7}^{max}$ \eqref{critical_fiducial}, the way each quantity depends on the free parameters of the model is different. It is this difference that can help us to fulfil our ``ideal model" more easily. 
Let us first consider the extension that is based upon the results of Schive et al.
In fact, using \eqref{m_hcrit_schive} and \eqref{critical_fiducial}, our ``ideal model" works out, as long as the SFDM parameters fulfil 
\begin{subequations}
\begin{equation}
    |\hat\Lambda| = 4.29\times 10^6 - 9.24\times 10^7,
\end{equation}
\begin{equation}
    m_{22}|\hat\Lambda|^{1/2} = 2.05\times 10^{4} - 2.05\times 10^{6}.
\end{equation}
This region of parameters is shown in figure \ref{fig:mass_selfneg_crit} (they are marked in green in the upper and lower figures). We can combine these two last expressions to obtain an estimate for the mass parameter. For example, if we adopt the value $|\hat\Lambda| = 4.29\times 10^6$ (i.e. if we demand to obtain collapse once $M_{h,12}^{(crit)} = 1$), the particle mass should be in the range $m_{22} = 9.90 - 9.90\times 10^2$ in order to fit our fiducial choice of  $M_{c,7}^{max} = 0.1-10$. If, on the other hand, we adopt $|\hat\Lambda| = 9.24\times 10^7$ (we demand to obtain collapse once $M_{h,12}^{(crit)} = 10^{-2}$), we obtain $m_{22} = 2.14-2.14\times 10^2$. Then, we can roughly estimate that the favoured mass for SMBH formation should be in the range
\begin{equation}\label{mass_estimate_schive}
    m_{22}\simeq 2.14- 9.90\times 10^2.
\end{equation}
\end{subequations}

Now, we turn to the extension which is based upon the results of Mocz et al., using \eqref{m_hcrit_mocz} and \eqref{critical_fiducial}. In this case, our ``ideal model" works out, as long as the SFDM parameters fulfil
\begin{subequations}
\begin{equation}
    m_{22}|\hat\Lambda|^{3/2} = 7.56\times 10^6 - 1.63\times 10^{10},
\end{equation}
\begin{equation}
    m_{22}|\hat\Lambda|^{1/2} = 2.05\times 10^{4} - 2.05\times 10^{6}.
\end{equation}
We have also shown this region of parameters in figure \ref{fig:mass_selfneg_crit} (they are marked in green in the middle and bottom figures). If we proceed to do a similar analysis than the one we did to estimate \eqref{mass_estimate_schive}, we finally arrive at the preferable parameters for SMBH formation:
\begin{equation}
    |\hat\Lambda| = 3.69 - 7.95 \times 10^5,
\end{equation}
\begin{equation}
    m_{22} = 22.99 - 1.7\times 10^4.
\end{equation}
\end{subequations}

Finally, we need to compare our estimates to previous constraints. This comparison is more difficult to do since, as we mentioned earlier, the self-interaction scenario has not been explored with much detail in this context. However, we can see that our estimates of the self-interaction term seem to be quite close to those reported by other works. Thus, in this part we decided to adopt our estimates as independent constraints for the possible formation of SMBHs. On the other hand, the region of parameters $|\hat\Lambda|>9.24\times 10^7$ for the Schive et al. extension, or $m_{22}|\hat\Lambda|^{3/2}> 1.63\times 10^{10}$ for our Mocz et al. extension, respectively, should be ruled out in order to avoid to form SMBHs for the lightest galactic halos (e.g. those harboring small dSphs). The same happens for the range of parameters $|\hat\Lambda| = 1.99\times 10^5-4.29\times 10^6$ and $m_{22}|\hat\Lambda|^{3/2} = 3.51\times 10^3-7.56\times 10^6$, which is ruled out in order to avoid collapse in galactic halos which are too massive. Finally, the \textit{non-formation} of SMBHs impose the constraint $|\hat\Lambda|\leq 1.99\times 10^5$, using our Schive et al. extension, and $m_{22}|\hat\Lambda|^{3/2}\leq 3.51\times 10^3$, using our Mocz et al. extension, respectively, which is a region of parameters {that fits previous constraints}.

\section{Conclusions}
\appendix

{We have studied the SFDM core-halo mass relations, which have been reported in various previous simulation papers. Our main objective was the extension of these relations for SFDM models which include self-interaction. }
After presenting the basic equations used to model SFDM, we adopted a Gaussian ansatz to describe typical core/soliton structures presented in this model. We showed that this ansatz can correctly reproduce several of the numerical results that are well known for these solitons, {with a special emphasis on the question beyond which critical mass those solitons will collapse.  This question is of immediate importance, once we realize that the core-halo mass relations imply that such soliton collapse could happen, once the mass of the halo itself exceeds a certain threshold. This has implications for SFDM with or without self-interaction.} We showed how the core-halo mass relation, typically found in numerical simulations of structure formation in the free SFDM model, can be generalized to models with self-interaction. Basically two different core-halo mass relations have been reported in the literature for free SFDM, hence we decided to extend both of them. 
{Using our extendend core-halo mass relations, we constrain the free parameters of the SFDM model by exploring the possibility of SMBH formation in the most massive galactic halos. Comparing our findings with previous constraints  reported by other groups through different observational evidence -- not related to the SMBH formation considered here --, we show that soliton collapse to form SMBHs is neither favoured in SFDM models without self-interaction, nor in those with repulsive self-interaction.
In these cases, the central solitons will never get close to the critical mass of collapse. However, if on the other hand we accept a range of parameters that are beyond those commonly reported for these two scenarios, i.e. if we adopted a much smaller de Broglie wave length for the free-field SFDM model or a much smaller (n=1)-polytrope radius for the repulsive scenario, it turns out that it is possible to explain the formation of SMBHs seeds with masses in the desired range for one of the two core-halo mass relations we explored. However, in adopting such a range, SFDM becomes indistinguishable from CDM during structure formation, since
the scale of suppressing small-scale structure is greatly reduced for that range of parameters. However, this does not mean that such a
model of SFDM is not a viable one for cosmic DM, since it would be hard
to distinguish it from CDM (except perhaps by direct detection or annihilation
effects of the latter), and that additionally it would have a natural mechanism to explain the formation of SMBHs. To conclude with our study, we found that in SFDM with attractive self-interaction, SMBH formation is feasible more easily since it is possible to fulfil our ``ideal model" of SMBH formation completely and for both core-halo mass extensions. Since not many studies have been done that constrain the free parameters of this scenario and that do not ignore the contribution of the self-interaction term (usually, in the attractive scenario it is assumed that the value of the self-interaction parameter must be extremely small, and then, this scenario is modelled effectively as a free-field, i.e. it is usually assumed that most of the constraints that are found in the free-field model are also shared for this scenario), we decided to elevate our results, that we obtained for this case, as independent constraints for the model.}
 
{More simulation work will be required in order to settle the question of which core-halo mass relation should be expected in SFDM models with and without self-interaction. In this work, we have built upon the existing literature which presents us with two different exponents for the core-halo mass relation in free SFDM models. We have taken them at face value, performing analytic calculations in order to show for each case which modified relations are expected, once self-interaction is included. This way, our work makes clean predictions which can be compared to upcoming simulations of SFDM structure formation. } 

\begin{acknowledgements}
L.P. acknowledge financial support from CONACyT doctoral fellowship. JAV acknowledges the support provided by  SEP-CONACYT Investigaci\'on B\'asica A1-S-21925, UNAM-PAPIIT  IA102219 and Ciencia de Frontera FORDECYT-PRONACES/304001/2020.
TRD is supported by the Austrian Science Fund FWF through an Elise
Richter fellowship (grant nr. V 656-N28).
This work was partially supported by CONACyT M\'exico under grants CB-2017 A1-S-8742, CB-2014-01 No. 240512, CB-2017 A1-S-17899, No. 269652, 304001, 376127;
Xiuhcoatl and Abacus clusters at Cinvestav, IPN;
I0101/131/07 C-234/07 of the Instituto
Avanzado de Cosmolog\'ia (IAC) collaboration (http://www.iac.edu.mx/). This research received support by Conacyt through the 
Fondo Sectorial de Investigaci\'on para la Educaci\'on, grant No. 240512 (TM).
\end{acknowledgements}

\section{Analytic approximations: Gaussian vs. rational function}\label{ap:gvsp}

Previous literature has made extensive use of two different analytic approaches for the central soliton in SFDM halos without self-interaction (the free case). On the one hand, there is a rational function density distribution $\rho_c^{(p)}$ given by \citep{sc4} 
\begin{equation}\label{denp}
    \rho_c^{(p)}(r) =  \frac{\rho_0}{\left(1+0.091\left(\frac{r}{r_c}\right)^2\right)^8},
\end{equation}
where $\rho_0$ is the central density of the soliton
\begin{equation}
    \rho_0 = 1.93\times 10^7 m_{22}^{-2}\left(\frac{r_c}{1kpc}\right)^{-4}M_\odot kpc^{-3},
\end{equation}
and the core radius $r_c$ is defined as the radius where the mass density drops by a factor of 2 from its value at the origin
\begin{equation}\label{r_c}
    r_c \simeq \frac{2.27\times 10^{4}}{(m_{22})^2M_{c,7}}pc.
\end{equation}
On the other hand, it has been noted that the soliton profile can be well approximated by a Gaussian density distribution $\rho_c^{(g)}$ \citep{sc15}
\begin{equation}\label{deng}
    \rho_c^{(g)}(r)= \frac{M_c}{(\pi R_c^2)^{3/2}}e^{-r^2/R_c^2},
\end{equation}
where we take $R_c$ in such a way that the radius that contains $99\%$ of the mass of the Gaussian ansatz matches with the numerical solution. Then,
\begin{equation}\label{R_c}
R_c \simeq \frac{3.54\times 10^4}{(m_{22})^2 M_{c,7}}pc.
\end{equation}
Observe from \eqref{tilde} that both cases, \eqref{denp} and \eqref{deng}, follow the same re-scaling dependence $\rho_c^{(p)},\rho_c^{(g)}\propto \gamma^{-4}$, as expected.

\begin{figure}[t!]
    \centering
    \includegraphics[width=3.4in]{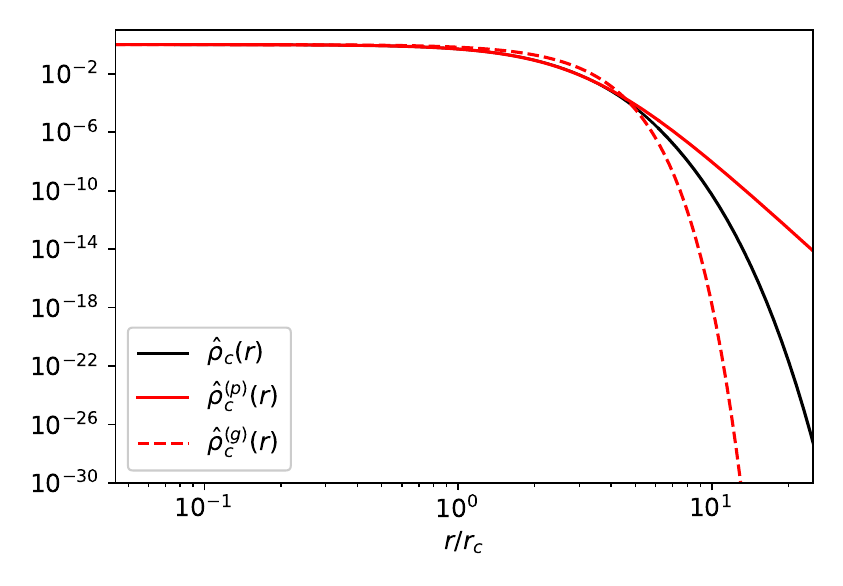}
        \includegraphics[width=3.4in]{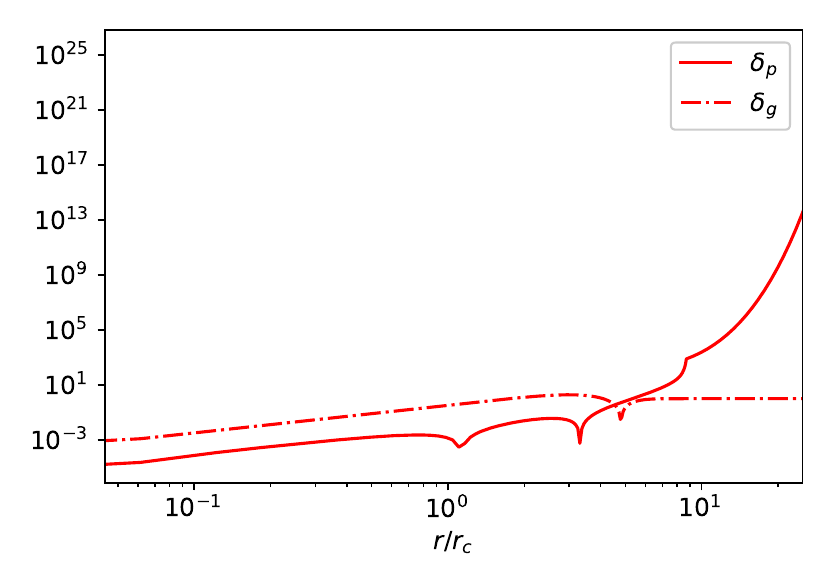}
        
      \includegraphics[width=3.4in]{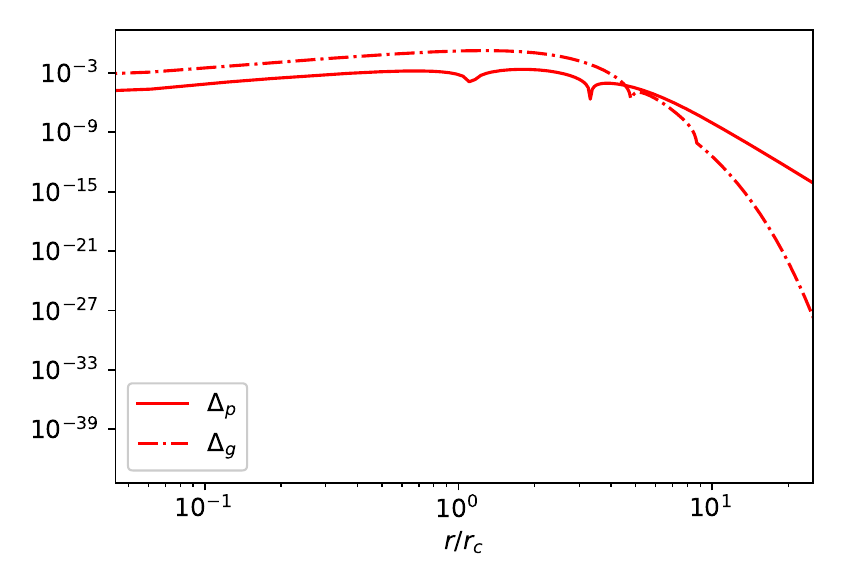}
    \caption{\footnotesize{Rational function (red solid) vs. Gaussian (red dashed) density distributions. In the top figure, we plot each case and the numerical solution (black solid), while in the bottom figures we plot the relative (middle) and the absolute (bottom) errors for each approximation.}}
    \label{fig:sol_comp}
\end{figure}

We can compare the above analytic profiles with the numerical solution. For that purpose, it is convenient to rewrite each approximation in terms of dimensionless variables \eqref{parameters}, i.e. ``hat" quantities, and by considering the solution that has a central scalar field value equal to 1. In this manner, we can compare each approximation with the numerical solution with $\gamma=1$. We notice that the analytic approach given in \eqref{denp} results in a better approximation for the soliton at small $\hat r$ than the Gaussian, as can be seen from figure \ref{fig:sol_comp}. In the top figure, we plot the dimensionless squared wave solution $|\hat\psi^{(1)}|^2$, where superscript 1 refers to $\gamma=1$, together with the Gaussian and the rational function approximations. The middle figure shows the relative error $\delta_i \equiv |(\hat\rho_c^{(1)}-\hat\rho_c^{(i)})/\hat\rho_c^{(i)}|$, $i=p,g$, while the bottom figure shows the total error $\Delta_i \equiv |\hat\rho_c^{(1)}-\hat\rho_c^{(i)}|$, $i=p,g$ for each approximation. 

\newpage
\bibliographystyle{ieeetr}
\bibliography{references}

\begin{thebibliography}{100}

\bibitem{cdm1}
P.~Peebles, ``Large-scale background temperature and mass fluctuations due to
  scale-invariant primeval perturbations,'' 1982.

\bibitem{cdm2}
S.~D. White, C.~S. Frenk, M.~Davis, and G.~Efstathiou, ``Clusters, filaments,
  and voids in a universe dominated by cold dark matter,'' {\em The
  Astrophysical Journal}, vol.~313, pp.~505--516, 1987.

\bibitem{cdm3}
J.~S. Bullock and M.~Boylan-Kolchin, ``Small-scale challenges to the
  $\lambda$cdm paradigm,'' {\em Annual Review of Astronomy and Astrophysics},
  vol.~55, 2017.

\bibitem{cdm4}
D.~Clowe, M.~Brada{\v{c}}, A.~H. Gonz\'alez, M.~Markevitch, S.~W. Randall,
  C.~Jones, and D.~Zaritsky, ``A direct empirical proof of the existence of
  dark matter,'' {\em The Astrophysical Journal Letters}, vol.~648, no.~2,
  p.~L109, 2006.

\bibitem{cdm5}
A.~Klypin, A.~V. Kravtsov, O.~Valenzuela, and F.~Prada, ``Where are the missing
  galactic satellites?,'' {\em The Astrophysical Journal}, vol.~522, no.~1,
  p.~82, 1999.

\bibitem{cdm6}
B.~Moore, S.~Ghigna, F.~Governato, and G.~Lake, ``G., t. quinn, j. stadel, \&
  p. tozzi,'' {\em ApJ Letters}, vol.~524, p.~L19, 1999.

\bibitem{cdm7}
S.~J. Penny, C.~J. Conselice, S.~De~Rijcke, and E.~V. Held, ``Hubble space
  telescope survey of the perseus cluster--i. the structure and dark matter
  content of cluster dwarf spheroidals,'' {\em Monthly Notices of the Royal
  Astronomical Society}, vol.~393, no.~3, pp.~1054--1062, 2009.

\bibitem{cdm8}
J.~M. Gaskins, ``A review of indirect searches for particle dark matter,'' {\em
  Contemporary Physics}, vol.~57, no.~4, pp.~496--525, 2016.

\bibitem{sf5}
A.~Arbey, J.~Lesgourgues, and P.~Salati, ``Cosmological constraints on
  quintessential halos,'' {\em Physical Review D}, vol.~65, no.~8, p.~083514,
  2002.

\bibitem{sf1}
V.~Sahni and L.~Wang, ``New cosmological model of quintessence and dark
  matter,'' {\em Physical Review D}, vol.~62, no.~10, p.~103517, 2000.

\bibitem{sf2}
W.~Hu, R.~Barkana, and A.~Gruzinov, ``Fuzzy cold dark matter: the wave
  properties of ultralight particles,'' {\em Physical Review Letters}, vol.~85,
  no.~6, p.~1158, 2000.

\bibitem{sf3}
T.~Matos and L.~A. Ure\~na L\'opez, ``Further analysis of a cosmological model
  with quintessence and scalar dark matter,'' {\em Physical Review D}, vol.~63,
  no.~6, p.~063506, 2001.

\bibitem{sf40}
T.~Matos and L.~A. Ure\~na L\'opez, ``Quintessence and scalar dark matter in
  the universe,'' {\em Classical and Quantum Gravity}, vol.~17, no.~13, p.~L75,
  2000.

\bibitem{sf4}
T.~Matos, F.~S. Guzm{\'a}n, and L.~A. Ure\~na L{\'o}pez, ``Scalar field as dark
  matter in the universe,'' {\em Classical and Quantum Gravity}, vol.~17,
  no.~7, p.~1707, 2000.

\bibitem{sf6}
A.~Arbey, J.~Lesgourgues, and P.~Salati, ``Quintessential halos around
  galaxies,'' {\em Physical Review D}, vol.~64, no.~12, p.~123528, 2001.

\bibitem{sf7}
A.~Arbey, J.~Lesgourgues, and P.~Salati, ``Galactic halos of fluid dark
  matter,'' {\em Physical Review D}, vol.~68, no.~2, p.~023511, 2003.

\bibitem{sf80}
S.-J. Sin, ``Late-time phase transition and the galactic halo as a bose
  liquid,'' {\em Physical Review D}, vol.~50, no.~6, p.~3650, 1994.

\bibitem{sf8}
S.~Ji and S.-J. Sin, ``Late-time phase transition and the galactic halo as a
  bose liquid. ii. the effect of visible matter,'' {\em Physical Review D},
  vol.~50, no.~6, p.~3655, 1994.

\bibitem{sc4}
H.-Y. Schive, T.~Chiueh, and T.~Broadhurst, ``Cosmic structure as the quantum
  interference of a coherent dark wave,'' {\em Nature Physics}, vol.~10, no.~7,
  p.~496, 2014.

\bibitem{sf10}
C.~Boehmer and T.~Harko, ``Can dark matter be a bose--einstein condensate?,''
  {\em Journal of Cosmology and Astroparticle Physics}, vol.~2007, no.~06,
  p.~025, 2007.

\bibitem{sf11}
D.~J. Marsh and P.~G. Ferreira, ``Ultralight scalar fields and the growth of
  structure in the universe,'' {\em Physical Review D}, vol.~82, no.~10,
  p.~103528, 2010.

\bibitem{sf12}
M.~Membrado, A.~Pacheco, and J.~Sa{\~n}udo, ``Hartree solutions for the
  self-yukawian boson sphere,'' {\em Physical Review A}, vol.~39, no.~8,
  p.~4207, 1989.

\bibitem{sf13}
F.~S. Guzm\'an, T.~Matos, and H.~Villegas, ``Scalar fields as dark matter in
  spiral galaxies: comparison with experiments,'' {\em Astronomische
  Nachrichten: News in Astronomy and Astrophysics}, vol.~320, no.~3,
  pp.~97--104, 1999.

\bibitem{sf14}
F.~S. Guzm{\'a}n and T.~Matos, ``Scalar fields as dark matter in spiral
  galaxies,'' {\em Classical and Quantum Gravity}, vol.~17, no.~1, p.~L9, 2000.

\bibitem{rev1}
J.~Maga\~na and T.~Matos, ``A brief review of the scalar field dark matter
  model,'' in {\em Journal of Physics: Conference Series}, vol.~378, p.~012012,
  IOP Publishing, 2012.

\bibitem{rev2}
A.~Su{\'a}rez, V.~H. Robles, and T.~Matos, ``A review on the scalar
  field/bose-einstein condensate dark matter model,'' in {\em Accelerated
  Cosmic Expansion}, pp.~107--142, Springer, 2014.

\bibitem{rev3}
D.~J. Marsh, ``Axion cosmology,'' {\em Physics Reports}, vol.~643, pp.~1--79,
  2016.

\bibitem{rev4}
L.~Hui, J.~P. Ostriker, S.~Tremaine, and E.~Witten, ``Ultralight scalars as
  cosmological dark matter,'' {\em Physical Review D}, vol.~95, no.~4,
  p.~043541, 2017.

\bibitem{niemeyer2019small}
J.~C. Niemeyer, ``Small-scale structure of fuzzy and axion-like dark matter,''
  {\em arXiv preprint arXiv:1912.07064}, 2019.

\bibitem{RS}
T.~Rindler-Daller and P.~R. Shapiro, ``Complex scalar field dark matter on
  galactic scales,'' {\em Modern Physics Letters A}, vol.~29, no.~02,
  p.~1430002, 2014.

\bibitem{sc10}
H.-Y. Schive, M.-H. Liao, T.-P. Woo, S.-K. Wong, T.~Chiueh, T.~Broadhurst, and
  W.~P. Hwang, ``Understanding the core-halo relation of quantum wave dark
  matter from 3d simulations,'' {\em Physical review letters}, vol.~113,
  no.~26, p.~261302, 2014.

\bibitem{sc11}
B.~Schwabe, J.~C. Niemeyer, and J.~F. Engels, ``Simulations of solitonic core
  mergers in ultralight axion dark matter cosmologies,'' {\em Physical Review
  D}, vol.~94, no.~4, p.~043513, 2016.

\bibitem{sc12}
J.~Veltmaat and J.~C. Niemeyer, ``Cosmological particle-in-cell simulations
  with ultralight axion dark matter,'' {\em Physical Review D}, vol.~94,
  no.~12, p.~123523, 2016.

\bibitem{sc13}
P.~Mocz, M.~Vogelsberger, V.~H. Robles, J.~Zavala, M.~Boylan-Kolchin,
  A.~Fialkov, and L.~Hernquist, ``Galaxy formation with becdm--i. turbulence
  and relaxation of idealized haloes,'' {\em Monthly Notices of the Royal
  Astronomical Society}, vol.~471, no.~4, pp.~4559--4570, 2017.

\bibitem{sc14}
D.~Levkov, A.~Panin, and I.~Tkachev, ``Gravitational bose-einstein condensation
  in the kinetic regime,'' {\em Physical review letters}, vol.~121, no.~15,
  p.~151301, 2018.

\bibitem{moczext1}
M.~Mina, D.~F. Mota, and H.~A. Winther, ``Solitons in the dark: non-linear
  structure formation with fuzzy dark matter,'' {\em arXiv preprint
  arXiv:2007.04119}, 2020.

\bibitem{moczext2}
M.~Nori and M.~Baldi, ``Scaling relations of fuzzy dark matter haloes i:
  individual systems in their cosmological environment,'' {\em arXiv preprint
  arXiv:2007.01316}, 2020.

\bibitem{sc15}
P.-H. Chavanis, ``Mass-radius relation of newtonian self-gravitating
  bose-einstein condensates with short-range interactions. i. analytical
  results,'' {\em Physical Review D}, vol.~84, no.~4, p.~043531, 2011.

\bibitem{sc16}
D.~J. Marsh and A.-R. Pop, ``Axion dark matter, solitons and the cusp--core
  problem,'' {\em Monthly Notices of the Royal Astronomical Society}, vol.~451,
  no.~3, pp.~2479--2492, 2015.

\bibitem{sc17}
S.-R. Chen, H.-Y. Schive, and T.~Chiueh, ``Jeans analysis for dwarf spheroidal
  galaxies in wave dark matter,'' {\em Monthly Notices of the Royal
  Astronomical Society}, vol.~468, no.~2, pp.~1338--1348, 2017.

\bibitem{sm2}
M.~Cappellari, ``Astrophysics: Monster black holes,'' {\em Nature}, vol.~480,
  no.~7376, p.~187, 2011.

\bibitem{sm3}
N.~J. McConnell, C.-P. Ma, K.~Gebhardt, S.~A. Wright, J.~D. Murphy, T.~R.
  Lauer, J.~R. Graham, and D.~O. Richstone, ``Two ten-billion-solar-mass black
  holes at the centres of giant elliptical galaxies,'' {\em Nature}, vol.~480,
  no.~7376, p.~215, 2011.

\bibitem{smbh1}
Y.~Matsuoka, M.~Onoue, N.~Kashikawa, M.~A. Strauss, K.~Iwasawa, C.-H. Lee,
  M.~Imanishi, T.~Nagao, M.~Akiyama, N.~Asami, {\em et~al.}, ``Discovery of the
  first low-luminosity quasar at z> 7,'' {\em The Astrophysical Journal
  Letters}, vol.~872, no.~1, p.~L2, 2019.

\bibitem{smbh2}
Y.~Matsuoka, M.~A. Strauss, N.~Kashikawa, M.~Onoue, K.~Iwasawa, J.-J. Tang,
  C.-H. Lee, M.~Imanishi, T.~Nagao, M.~Akiyama, {\em et~al.}, ``Subaru high-z
  exploration of low-luminosity quasars (shellqs). v. quasar luminosity
  function and contribution to cosmic reionization at z= 6,'' {\em The
  Astrophysical Journal}, vol.~869, no.~2, p.~150, 2018.

\bibitem{smbh3}
Y.~Matsuoka, K.~Iwasawa, M.~Onoue, N.~Kashikawa, M.~A. Strauss, C.-H. Lee,
  M.~Imanishi, T.~Nagao, M.~Akiyama, N.~Asami, {\em et~al.}, ``Subaru high-z
  exploration of low-luminosity quasars (shellqs). iv. discovery of 41 quasars
  and luminous galaxies at $5.7\leq z\leq 6.9$,'' {\em The Astrophysical
  Journal Supplement Series}, vol.~237, no.~1, p.~5, 2018.

\bibitem{smbh4}
Y.~Matsuoka, M.~Onoue, N.~Kashikawa, K.~Iwasawa, M.~A. Strauss, T.~Nagao,
  M.~Imanishi, C.-H. Lee, M.~Akiyama, N.~Asami, {\em et~al.}, ``Subaru high-z
  exploration of low-luminosity quasars (shellqs). ii. discovery of 32 quasars
  and luminous galaxies at $5.7< z\leq 6.8$,'' {\em Publications of the
  Astronomical Society of Japan}, vol.~70, no.~SP1, p.~S35, 2017.

\bibitem{smbh5}
Y.~Matsuoka, M.~Onoue, N.~Kashikawa, K.~Iwasawa, M.~A. Strauss, T.~Nagao,
  M.~Imanishi, M.~Niida, Y.~Toba, M.~Akiyama, {\em et~al.}, ``Subaru high-z
  exploration of low-luminosity quasars (shellqs). i. discovery of 15 quasars
  and bright galaxies at 5.7< z< 6.9,'' {\em The Astrophysical Journal},
  vol.~828, no.~1, p.~26, 2016.

\bibitem{smbh6}
X.-B. Wu, F.~Wang, X.~Fan, W.~Yi, W.~Zuo, F.~Bian, L.~Jiang, I.~D. McGreer,
  R.~Wang, J.~Yang, {\em et~al.}, ``An ultraluminous quasar with a
  twelve-billion-solar-mass black hole at redshift 6.30,'' {\em Nature},
  vol.~518, no.~7540, p.~512, 2015.

\bibitem{smbh7}
X.~Fan, M.~A. Strauss, D.~P. Schneider, R.~H. Becker, R.~L. White, Z.~Haiman,
  M.~Gregg, L.~Pentericci, E.~K. Grebel, V.~K. Narayanan, {\em et~al.}, ``A
  survey of z> 5.7 quasars in the sloan digital sky survey. ii. discovery of
  three additional quasars at z> 6,'' {\em The Astronomical Journal}, vol.~125,
  no.~4, p.~1649, 2003.

\bibitem{smbh8}
L.~Jiang, X.~Fan, M.~Vestergaard, J.~D. Kurk, F.~Walter, B.~C. Kelly, and M.~A.
  Strauss, ``Gemini near-infrared spectroscopy of luminous z~ 6 quasars:
  chemical abundances, black hole masses, and mg ii absorption,'' {\em The
  Astronomical Journal}, vol.~134, no.~3, p.~1150, 2007.

\bibitem{smbh9}
C.~J. Willott, P.~Delorme, A.~Omont, J.~Bergeron, X.~Delfosse, T.~Forveille,
  L.~Albert, C.~Reyl{\'e}, G.~J. Hill, M.~Gully-Santiago, {\em et~al.}, ``Four
  quasars above redshift 6 discovered by the canada-france high-z quasar
  survey,'' {\em The Astronomical Journal}, vol.~134, no.~6, p.~2435, 2007.

\bibitem{smbh10}
L.~Jiang, X.~Fan, J.~Annis, R.~H. Becker, R.~L. White, K.~Chiu, H.~Lin, R.~H.
  Lupton, G.~T. Richards, M.~A. Strauss, {\em et~al.}, ``A survey of z~ 6
  quasars in the sloan digital sky survey deep stripe. i. a flux-limited sample
  at zab< 21,'' {\em The Astronomical Journal}, vol.~135, no.~3, p.~1057, 2008.

\bibitem{smbh11}
C.~J. Willott, P.~Delorme, C.~Reyl{\'e}, L.~Albert, J.~Bergeron, D.~Crampton,
  X.~Delfosse, T.~Forveille, J.~B. Hutchings, R.~J. McLure, {\em et~al.}, ``The
  canada-france high-z quasar survey: Nine new quasars and the luminosity
  function at redshift 6,'' {\em The Astronomical Journal}, vol.~139, no.~3,
  p.~906, 2010.

\bibitem{smbh12}
D.~J. Mortlock, S.~J. Warren, B.~P. Venemans, M.~Patel, P.~C. Hewett, R.~G.
  McMahon, C.~Simpson, T.~Theuns, E.~A. Gonz{\'a}les-Solares, A.~Adamson, {\em
  et~al.}, ``A luminous quasar at a redshift of z= 7.085,'' {\em Nature},
  vol.~474, no.~7353, p.~616, 2011.

\bibitem{smbh13}
B.~Venemans, J.~Findlay, W.~Sutherland, G.~De~Rosa, R.~McMahon, R.~Simcoe,
  E.~Gonz{\'a}lez-Solares, K.~Kuijken, and J.~Lewis, ``Discovery of three z>
  6.5 quasars in the vista kilo-degree infrared galaxy (viking) survey,'' {\em
  The Astrophysical Journal}, vol.~779, no.~1, p.~24, 2013.

\bibitem{smbh14}
E.~Ba{\~n}ados, B.~Venemans, E.~Morganson, R.~Decarli, F.~Walter, K.~Chambers,
  H.-W. Rix, E.~Farina, X.~Fan, L.~Jiang, {\em et~al.}, ``Discovery of eight z~
  6 quasars from pan-starrs1,'' {\em The Astronomical Journal}, vol.~148,
  no.~1, p.~14, 2014.

\bibitem{cloud_c}
J.~Silk and M.~J. Rees, ``Quasars and galaxy formation,'' {\em arXiv preprint
  astro-ph/9801013}, 1998.

\bibitem{KF}
K.~{Freese}, C.~{Ilie}, D.~{Spolyar}, M.~{Valluri}, and P.~{Bodenheimer},
  ``{Supermassive Dark Stars: Detectable in JWST},'' {\em \apj}, vol.~716,
  pp.~1397--1407, Jun 2010.

\bibitem{R15}
T.~{Rindler-Daller}, M.~H. {Montgomery}, K.~{Freese}, D.~E. {Winget}, and
  B.~{Paxton}, ``{Dark Stars: Improved Models and First Pulsation Results},''
  {\em \apj}, vol.~799, p.~210, Feb 2015.

\bibitem{merger}
K.~Menou, Z.~Haiman, and V.~K. Narayanan, ``The merger history of supermassive
  black holes in galaxies,'' {\em The Astrophysical Journal}, vol.~558, no.~2,
  p.~535, 2001.

\bibitem{correlation1}
D.~Merritt and L.~Ferrarese, ``The m•-$\sigma$ relation for supermassive
  black holes,'' {\em The Astrophysical Journal}, vol.~547, no.~1, p.~140,
  2001.

\bibitem{correlation2}
K.~Gebhardt, R.~Bender, G.~Bower, A.~Dressler, S.~Faber, A.~V. Filippenko,
  R.~Green, C.~Grillmair, L.~C. Ho, J.~Kormendy, {\em et~al.}, ``A relationship
  between nuclear black hole mass and galaxy velocity dispersion,'' {\em The
  Astrophysical Journal Letters}, vol.~539, no.~1, p.~L13, 2000.

\bibitem{smbhcorr}
L.~Ferrarese, ``Beyond the bulge: a fundamental relation between supermassive
  black holes and dark matter halos,'' {\em The Astrophysical Journal},
  vol.~578, no.~1, p.~90, 2002.

\bibitem{smbhcorr2}
K.~Bandara, D.~Crampton, and L.~Simard, ``A relationship between supermassive
  black hole mass and the total gravitational mass of the host galaxy,'' {\em
  The Astrophysical Journal}, vol.~704, no.~2, p.~1135, 2009.

\bibitem{sm7}
A.~\'Avilez, T.~Bernal, L.~E. Padilla, and T.~Matos, ``On the possibility that
  ultra-light boson haloes host and form supermassive black holes,'' {\em
  Monthly Notices of the Royal Astronomical Society}, vol.~477, no.~3,
  pp.~3257--3272, 2018.

\bibitem{boson_stars1}
E.~Seidel and W.-M. Suen, ``Dynamical evolution of boson stars: Perturbing the
  ground state,'' {\em Physical Review D}, vol.~42, no.~2, p.~384, 1990.

\bibitem{sm9}
E.~Seidel and W.-M. Suen, ``Oscillating soliton stars,'' {\em Physical review
  letters}, vol.~66, no.~13, p.~1659, 1991.

\bibitem{sm10}
E.~Seidel and W.-M. Suen, ``Formation of solitonic stars through gravitational
  cooling,'' {\em Physical review letters}, vol.~72, no.~16, p.~2516, 1994.

\bibitem{sm11}
N.~Sanchis-Gual, J.~C. Degollado, J.~A. Font, C.~Herdeiro, and E.~Radu,
  ``Dynamical formation of a hairy black hole in a cavity from the decay of
  unstable solitons,'' {\em Classical and Quantum Gravity}, vol.~34, no.~16,
  p.~165001, 2017.

\bibitem{sm12}
M.~Alcubierre, F.~S. Guzm{\'a}n, T.~Matos, D.~Nu\~nez, L.~A. Ure\~na L\'opez,
  and P.~Wiederhold, ``Galactic collapse of scalar field dark matter,'' {\em
  Classical and Quantum Gravity}, vol.~19, no.~19, p.~5017, 2002.

\bibitem{sm15}
J.~Barranco, A.~Bernal, J.~C. Degollado, A.~Diez-Tejedor, M.~Megevand,
  M.~Alcubierre, D.~N{\'u}{\~n}ez, and O.~Sarbach, ``Are black holes a serious
  threat to scalar field dark matter models?,'' {\em Physical Review D},
  vol.~84, no.~8, p.~083008, 2011.

\bibitem{sm16}
J.~Barranco, A.~Bernal, J.~C. Degollado, A.~Diez-Tejedor, M.~Megevand,
  M.~Alcubierre, D.~N{\'u}{\~n}ez, and O.~Sarbach, ``Schwarzschild black holes
  can wear scalar wigs,'' {\em Physical review letters}, vol.~109, no.~8,
  p.~081102, 2012.

\bibitem{wf1}
R.~Ruffini and S.~Bonazzola, ``Systems of self-gravitating particles in general
  relativity and the concept of an equation of state,'' {\em Physical Review},
  vol.~187, no.~5, p.~1767, 1969.

\bibitem{ure-guz}
F.~S. Guzm\'an and L.~A. Urena-L{\'o}pez, ``Evolution of the
  schr{\"o}dinger-newton system for a self-gravitating scalar field,'' {\em
  Physical Review D}, vol.~69, no.~12, p.~124033, 2004.

\bibitem{wf3}
F.~S. Guzm\'an and L.~A. Ure\~na L\'opez, ``Gravitational cooling of
  self-gravitating bose condensates,'' {\em The Astrophysical Journal},
  vol.~645, no.~2, p.~814, 2006.

\bibitem{wf2}
N.~Bar, D.~Blas, K.~Blum, and S.~Sibiryakov, ``Galactic rotation curves versus
  ultralight dark matter: Implications of the soliton-host halo relation,''
  {\em Physical Review D}, vol.~98, no.~8, p.~083027, 2018.

\bibitem{core_halo1}
N.~Bar, K.~Blum, R.~Sato, and J.~Eby, ``arxiv: Ultra-light dark matter in disk
  galaxies,'' tech. rep., 2019.

\bibitem{boson_stars}
J.~Eby, C.~Kouvaris, N.~G. Nielsen, and L.~Wijewardhana, ``Boson stars from
  self-interacting dark matter,'' {\em Journal of High Energy Physics},
  vol.~2016, no.~2, p.~28, 2016.

\bibitem{guzman2018head}
F.~Guzm{\'a}n and A.~A. \'Avilez, ``Head-on collision of multistate ultralight
  bec dark matter configurations,'' {\em Physical Review D}, vol.~97, no.~11,
  p.~116003, 2018.

\bibitem{BP}
G.~{Baym} and C.~J. {Pethick}, ``{Ground-State Properties of Magnetically
  Trapped Bose-Condensed Rubidium Gas},'' {\em \prl}, vol.~76, pp.~6--9, Jan
  1996.

\bibitem{goodman2000repulsive}
J.~Goodman, ``Repulsive dark matter,'' {\em New Astronomy}, vol.~5, no.~2,
  pp.~103--107, 2000.

\bibitem{peebles2000fluid}
P.~Peebles, ``Fluid dark matter,'' {\em The Astrophysical Journal Letters},
  vol.~534, no.~2, p.~L127, 2000.

\bibitem{dynamical}
S.~L. Liebling and C.~Palenzuela, ``Dynamical boson stars,'' {\em Living
  Reviews in Relativity}, vol.~20, no.~1, p.~5, 2017.

\bibitem{colpi}
M.~Colpi, S.~L. Shapiro, and I.~Wasserman, ``Boson stars: Gravitational
  equilibria of self-interacting scalar fields,'' {\em Physical review
  letters}, vol.~57, no.~20, p.~2485, 1986.

\bibitem{chavanis2019core}
P.-H. Chavanis, ``Core mass--halo mass relation of bosonic and fermionic dark
  matter halos harbouring a supermassive black hole,'' {\em arXiv preprint
  arXiv:1911.01937}, 2019.

\bibitem{chavanis2019predictive}
P.-H. Chavanis, ``Predictive model of bec dark matter halos with a solitonic
  core and an isothermal atmosphere,'' {\em Physical Review D}, vol.~100,
  no.~8, p.~083022, 2019.

\bibitem{chavanis2018derivation}
P.-H. Chavanis, ``Derivation of a generalized schr{\"o}dinger equation for dark
  matter halos from the theory of scale relativity,'' {\em Physics of the Dark
  Universe}, vol.~22, pp.~80--95, 2018.

\bibitem{ahn2017detection}
C.~P. Ahn, A.~C. Seth, M.~Den~Brok, J.~Strader, H.~Baumgardt, R.~Van Den~Bosch,
  I.~Chilingarian, M.~Frank, M.~Hilker, R.~McDermid, {\em et~al.}, ``Detection
  of supermassive black holes in two virgo ultracompact dwarf galaxies,'' {\em
  The Astrophysical Journal}, vol.~839, no.~2, p.~72, 2017.

\bibitem{dsphsmbh}
A.~E. Reines, J.~J. Condon, J.~Darling, and J.~E. Greene, ``A new sample of
  (wandering) massive black holes in dwarf galaxies from high-resolution radio
  observations,'' {\em The Astrophysical Journal}, vol.~888, no.~1, p.~36,
  2020.

\bibitem{abbott2020gw190521}
R.~Abbott, T.~Abbott, S.~Abraham, F.~Acernese, K.~Ackley, C.~Adams,
  R.~Adhikari, V.~Adya, C.~Affeldt, M.~Agathos, {\em et~al.}, ``Gw190521: A
  binary black hole merger with a total mass of $150 ~m_\odot$,'' {\em Physical
  review letters}, vol.~125, no.~10, p.~101102, 2020.

\bibitem{free_const1}
A.~Paredes and H.~Michinel, ``Interference of dark matter solitons and galactic
  offsets,'' {\em Physics of the Dark Universe}, vol.~12, pp.~50--55, 2016.

\bibitem{free_const2}
A.~X. Gonz{\'a}lez-Morales, D.~J. Marsh, J.~Pe{\~n}arrubia, and L.~A.
  Ure{\~n}a-L{\'o}pez, ``Unbiased constraints on ultralight axion mass from
  dwarf spheroidal galaxies,'' {\em Monthly Notices of the Royal Astronomical
  Society}, vol.~472, no.~2, pp.~1346--1360, 2017.

\bibitem{free_const3}
V.~Lora, J.~Magana, A.~Bernal, F.~S{\'a}nchez-Salcedo, and E.~Grebel, ``On the
  mass of ultra-light bosonic dark matter from galactic dynamics,'' {\em
  Journal of Cosmology and Astroparticle Physics}, vol.~2012, no.~02, p.~011,
  2012.

\bibitem{free_const4}
E.~Calabrese and D.~N. Spergel, ``Ultra-light dark matter in ultra-faint dwarf
  galaxies,'' {\em Monthly Notices of the Royal Astronomical Society},
  vol.~460, no.~4, pp.~4397--4402, 2016.

\bibitem{free_const5}
A.~Sarkar, R.~Mondal, S.~Das, S.~K. Sethi, S.~Bharadwaj, and D.~J. Marsh, ``The
  effects of the small-scale dm power on the cosmological neutral hydrogen (hi)
  distribution at high redshifts,'' {\em Journal of Cosmology and Astroparticle
  Physics}, vol.~2016, no.~04, p.~012, 2016.

\bibitem{free_const6}
E.~Armengaud, N.~Palanque-Delabrouille, C.~Y{\`e}che, D.~J. Marsh, and J.~Baur,
  ``Constraining the mass of light bosonic dark matter using sdss
  lyman-$\alpha$ forest,'' {\em Monthly Notices of the Royal Astronomical
  Society}, vol.~471, no.~4, pp.~4606--4614, 2017.

\bibitem{free_const7}
V.~Ir{\v{s}}i{\v{c}}, M.~Viel, M.~G. Haehnelt, J.~S. Bolton, and G.~D. Becker,
  ``First constraints on fuzzy dark matter from lyman-$\alpha$ forest data and
  hydrodynamical simulations,'' {\em Physical review letters}, vol.~119, no.~3,
  p.~031302, 2017.

\bibitem{self-const1}
B.~Li, T.~Rindler-Daller, and P.~R. Shapiro, ``Cosmological constraints on
  bose-einstein-condensed scalar field dark matter,'' {\em Physical Review D},
  vol.~89, no.~8, p.~083536, 2014.

\bibitem{cosmo_self}
A.~Su{\'a}rez and P.-H. Chavanis, ``Cosmological evolution of a complex scalar
  field with repulsive or attractive self-interaction,'' {\em Physical Review
  D}, vol.~95, no.~6, p.~063515, 2017.

\bibitem{self-const2}
L.~E. Padilla, J.~A. V{\'a}zquez, T.~Matos, and G.~Germ{\'a}n, ``Scalar field
  dark matter spectator during inflation: the effect of self-interaction,''
  {\em Journal of Cosmology and Astroparticle Physics}, vol.~2019, no.~05,
  p.~056, 2019.

\bibitem{self-const3}
B.~Li, P.~R. Shapiro, and T.~Rindler-Daller, ``Bose-einstein-condensed scalar
  field dark matter and the gravitational wave background from inflation: new
  cosmological constraints and its detectability by ligo,'' {\em Physical
  Review D}, vol.~96, no.~6, p.~063505, 2017.

\bibitem{chavanis}
A.~Su{\'a}rez and P.-H. Chavanis, ``Hydrodynamic representation of the
  klein-gordon-einstein equations in the weak field limit: General formalism
  and perturbations analysis,'' {\em Physical Review D}, vol.~92, no.~2,
  p.~023510, 2015.

\bibitem{axion_like}
L.~Visinelli, ``Light axion-like dark matter must be present during
  inflation,'' {\em Physical Review D}, vol.~96, no.~2, p.~023013, 2017.

\bibitem{lowf}
M.~Cicoli, M.~D. Goodsell, and A.~Ringwald, ``The type iib string axiverse and
  its low-energy phenomenology,'' {\em Journal of High Energy Physics},
  vol.~2012, no.~10, p.~146, 2012.

\bibitem{supermassive}
P.-H. Chavanis, ``Collapse of a self-gravitating bose-einstein condensate with
  attractive self-interaction,'' {\em Physical Review D}, vol.~94, no.~8,
  p.~083007, 2016.

\end{thebibliography}
\end{document}